\documentclass[letterpaper,onecolumn,allowtoday,accepted=2026-09-02,preprintnumbers,10pt]{quantumarticle}
\pdfoutput=1
\usepackage[numbers,sort&compress]{natbib}
\usepackage[utf8]{inputenc}
\usepackage[english]{babel}
\usepackage[T1]{fontenc}
\usepackage{textgreek}
\usepackage{braket}
\usepackage{placeins}
\usepackage{graphicx,subfigure}
\usepackage{graphicx,amsmath,amsfonts,amssymb,amsthm,xr}
\usepackage{orcidlink}
\usepackage{physics}
\usepackage{graphicx}% Include figure files
\usepackage{dcolumn}% Align table columns on decimal point
\usepackage{bm}% bold math
\usepackage[capitalise]{cleveref}
\usepackage{subcaption}

\usepackage{color}

\begin{document}

\title{Truncation uncertainties for accurate quantum simulations of lattice gauge theories}

\author{Anthony N.~Ciavarella \,\orcidlink{0000-0003-3918-4110}}
\email{anciavarella@lbl.gov}
\affiliation{Physics Division, Lawrence Berkeley National Laboratory, Berkeley, California 94720, USA}
\author{Siddharth Hariprakash \,\orcidlink{0009-0000-3524-4724}}
\email{siddharth\_hari@berkeley.edu}
\affiliation{Physics Division, Lawrence Berkeley National Laboratory, Berkeley, California 94720, USA}
\affiliation{Leinweber Institute for Theoretical Physics and Department of Physics, University of California, Berkeley, California 94720, USA}
\author{Jad C.~Halimeh \,\orcidlink{0000-0002-0659-7990}}
\email{jad.halimeh@physik.lmu.de}
\affiliation{Max Planck Institute of Quantum Optics, 85748 Garching, Germany}
\affiliation{Department of Physics and Arnold Sommerfeld Center for Theoretical Physics (ASC), Ludwig Maximilian University of Munich, 80333 Munich, Germany}
\affiliation{Munich Center for Quantum Science and Technology (MCQST), 80799 Munich, Germany}
\author{Christian W.~Bauer \,\orcidlink{0000-0001-9820-5810}}
\email{cwbauer@lbl.gov}
\affiliation{Physics Division, Lawrence Berkeley National Laboratory, Berkeley, California 94720, USA}
\affiliation{Leinweber Institute for Theoretical Physics and Department of Physics, University of California, Berkeley, California 94720, USA}

\date{\today}

\begin{abstract}
The encoding of lattice gauge theories onto quantum computers requires a discretization of the gauge field's Hilbert space on each link, which presents errors with respect to the Kogut--Susskind limit. In the electric basis, Hilbert space fragmentation has recently been shown to limit the excitation of large electric fields. Here, we leverage this to develop a formalism for estimating the size of truncation errors in the electric basis. Generically, the truncation error falls off as a factorial of the field truncation. Examples of this formalism are applied to the Schwinger model and a pure U($1$) lattice gauge theory. For reasonable choices of parameters, we improve on previous error estimates by a factor of $10^{306}$.
\end{abstract}

\maketitle

%\tableofcontents

\section{Introduction}
Quantum computers offer the ability to directly simulate the real-time dynamics of quantum field theories~\cite{Feynman:1981tf}. This is anticipated to enable the prediction of dynamical non-perturbative quantities from quantum chromodynamics (QCD)~\cite{Humble:2022vtm,Bauer:2022hpo,DiMeglio:2023nsa,Beck:2023xhh}. As a particular example, quantum simulations of lattice QCD will enable the computation of soft functions, relevant to understanding the production of QCD jets from high-energy particle collisions~\cite{Bauer:2021gup,Bauer:2025nzf}. It will also enable computations of the quark-gluon plasma viscosity~\cite{Cohen:2021imf,Turro:2024pxu} and inelastic scattering amplitudes for hadron collisions~\cite{Luscher:1985dn,Luscher:1986pf,Ciavarella:2020vqm,Briceno:2020rar,Briceno:2021aiw,Carrillo:2024chu,Burbano:2025pef}. In addition to their relevance to high-energy physics, quantum simulations of lattice gauge theories are anticipated to give insights into topological phases and quantum spin liquids, where they emerge as an effective description of the dynamics~\cite{Levin:2004mi,Savary:2016ksw,Trebst:2022hxi,Motome:2019orm}. They have also been shown to demonstrate multiple forms of ergodicity-breaking behavior~\cite{Ciavarella:2025zqf,Desaules:2023ghs,Desaules:2024cua,Jeyaretnam:2024tkj}.

The large potential for quantum simulation and the recent development of noisy quantum computers has led to a large amount of work exploring how to use these devices to perform simulations. Lattice gauge theories with a number of different gauge groups in $1+1D$ have been simulated on quantum computers~\cite{Martinez:2016yna,Klco:2018kyo,Gorg:2018xyc,Mil:2019pbt,Yang:2020yer,Zhou:2021kdl,Wang:2021xra,Nguyen:2021hyk,Su:2022glk,Wang:2022dpp,Zhang:2023hzr,Angelides:2023noe,Mildenberger:2022jqr,Charles:2023zbl,Mueller:2022xbg,De:2024smi,Davoudi:2024wyv,Guo:2024tnb,Farrell:2023fgd,Farrell:2024fit,Atas:2021ext,Atas:2022dqm,Than:2024zaj,Farrell:2022wyt,Farrell:2022vyh,Ciavarella:2023mfc,Ciavarella:2024lsp,Chernyshev:2025lil,Chen:2025zeh,Zhu:2024dvz,Gyawali:2024hrz,Schuhmacher:2025ehh,Davoudi:2025rdv}. Limited simulations have been performed in higher spatial dimensions~\cite{Klco:2019evd,Paulson:2020zjd,Ciavarella:2021nmj,Ciavarella:2021lel,ARahman:2022tkr,Mendicelli:2022ntz,Turro:2024pxu,Li:2024lrl,Halimeh:2024ref,Gupta:2024gnw,Crippa:2024hso,Kavaki:2025hcu,Kavaki:2024ijd}. This is due to gauge fields having continuous degrees of freedom and quantum computers having finite, discrete degrees of freedom. This challenge is not present in $1+1D$ since Gauss's law can be used to integrate out gauge fields, leaving only fermions, which have a finite Hilbert space. This leads to long-range interactions; however, the range of this interaction can be truncated due to the exponential decay of correlations in low-energy states~\cite{Farrell:2024fit}. Mapping the gauge fields in higher dimensions onto quantum computers requires a truncation of the gauge fields. For the standard Kogut--Susskind Hamiltonian, a number of approaches have been developed, including electric basis truncations~\cite{Byrnes:2005qx,Banuls:2017ena,Klco:2019evd,Zohar:2019ygc,Ciavarella:2021nmj,Lewis:2019wfx,Raychowdhury:2018osk,Raychowdhury:2019iki,Stryker:2020vls,Kadam:2022ipf,Kadam:2023gpn,Kadam:2024zkj,Zache:2023dko,Kan:2021nyu,Ciavarella:2022zhe,Ciavarella:2024fzw,Muller:2023nnk,Kavaki:2024ijd,Hayata:2023bgh,Rigobello:2023ype,Fontana:2024rux,Balaji:2025afl,Illa:2025dou,Ciavarella:2025bsg}, discrete subgroups~\cite{Lamm:2019bik,Alexandru:2019nsa,Alam:2021uuq,Ji:2022qvr,Gustafson:2022xdt,Gustafson:2023swx,Gustafson:2023kvd,Gustafson:2024kym,Assi:2024pdn,Hartung:2022hoz}, and hybrid electric/magnetic bases~\cite{Bauer:2021gek,DAndrea:2023qnr,Jakobs:2023lpp,Garofalo:2023zkd,Grabowska:2024emw,Burbano:2024uvn,Jakobs:2025rvz}. Alternatives to the usual Kogut--Susskind Hamiltonian have also been developed, such as quantum link models~~\cite{Brower:1997ha,Brower:2003vy,Liu:2023lsr,Chandrasekharan:2025pil,Chandrasekharan:2025smw},  fuzzy gauge theories~\cite{Alexandru:2023qzd}, and orbifold lattice gauge theory~\cite{Buser:2020cvn,Bergner:2024qjl,Halimeh:2024bth,Lamm:2026zzl,Hanada:2026zab}, which are conjectured to have QCD as their continuum limit.

Ultimately, simulations capable of making predictions with precision will need to have controllable uncertainties. This will have contributions from both the hardware and the theoretical formulation of the simulation.  The effects of a finite lattice spacing can be estimated by matching to a continuum effective Symanzik action as is done in traditional lattice QCD~\cite{Symanzik:1983dc,Symanzik:1983gh,Sheikholeslami:1985ij,Illa:2025dou,Illa:2025njz} . This formalism can also be used to understand the error coming from discrete (possibly Trotterized) time steps~\cite{Carena:2021ltu,Kane:2025ybw}. Recent work has developed the formalism necessary for estimating finite volume errors in a generic correlation function computed on a quantum computer~\cite{Burbano:2025pef}. The truncation of the gauge fields will also contribute to the theoretical uncertainty. Several works have shown a fast convergence of energy eigenstates with field truncation for a number of different theories~\cite{Banuls:2017ena,Zache:2021ggw,Halimeh:2021ufh,Bruckmann:2018usp,Klco:2018zqz,Araz:2022tbd,Alexandru:2022son,Ciavarella:2022qdx,Ciavarella:2021nmj,Zache:2023dko,Yao:2023gnm}.  Additionally, one can further reduce truncation errors through adding perturbative corrections or by using similarity renormalization group methods~\cite{Ciavarella:2023mfc}. For scalar field theories, the exponential convergence is a reflection of the Nyquist-Shannon sampling theorem~\cite{Klco:2018zqz}, however a similar mechanism for non-Abelian gauge theories has not yet been identified. Previous work has shown that truncation errors in dynamics converge exponentially fast, provided one uses a truncation that grows linearly with time~\cite{Tong:2021rfv}. However, it is essential to have tight estimates of the truncation error, as overestimating errors will unnecessarily delay quantum simulations of scientific importance. In this work, the presence of Hilbert space fragmentation in the Kogut--Susskind Hamiltonian is used to obtain estimates of the truncation error for truncations in the electric basis. These error estimates do not require the truncation to grow with time to maintain accuracy and go to zero as a factorial in the size of the truncated link Hilbert space. Numerical simulations are performed for the Schwinger model and a pure U($1$) lattice gauge theory on a plaquette ladder. These simulations show that the error estimates in this work correctly capture the dynamics of lattice gauge theories.

\section{Truncation Errors in Pure Gauge Theories}
We begin this section by introducing the $U(1)$ lattice gauge theory on a single plaquette which will serve as a motivating example for the subsequent sections. The Hamiltonian is given by

\begin{align}
%\label{eq:HU1def}
    \hat{H} & = 2g^2 \hat{E}^2 + \frac{1}{2g^2} \left(2 - \hat{\Box} - \hat{\Box}^\dagger\right) \nonumber \\
\label{eq:EU1def}
    \hat{E} & = \sum_{n=-\infty}^{\infty} n \ket{n} \bra{n}  \nonumber \\
%\label{eq:BoxU1def}
    \hat{\Box} & = \sum_{n=-\infty}^{\infty} \ket{n} \bra{n+1}  \, ,
\end{align}
where $\hat{E}$ is the electric operator and the states $\ket{n}$ are the eigenstates of $\hat{E}$ that form the standard electric basis of the Kogut-Susskind Hamiltonian. $\hat{\Box} + \hat{\Box}^\dagger$ is the plaquette term which for a single plaquette connects electric basis states that differ by one unit of electric flux. We note that the methods in the following sections apply more generally beyond just the $U(1)$ theory as we will discuss. We will use $U(1)$ as a concrete example to demonstrate our methods.

\subsection{Eigenstate Truncation Errors}

To motivate why improvements in truncation error are expected to be possible, traditional perturbation theory will be applied to eigenstates of truncated Hamiltonians. We will consider a Hamiltonian, $\hat{H}$, defined on a single bosonic mode with basis states given by $\ket{0},\ket{1},\ket{2},\dots$ that are eigenstates of a number operator (similar to the electric basis states for the single plaquette $U(1)$ theory) and assume that the terms in the Hamiltonian do not change the boson number by more than $1$, i.e. $\bra{n+k}\hat{H}\ket{n}=0$ for $k>1$. For a lattice gauge theory, this would be analogous to a theory with a single plaquette and no matter. A truncated Hamiltonian, $\hat{H}_\Lambda$, can be defined by removing the off-diagonal elements of $\hat{H}$ above some truncation $\Lambda$. Explicitly, the relation between the truncated and untruncated Hamiltonian is given by
\begin{equation}
    \hat{H} = \hat{H}_\Lambda + \hat{V}_\Lambda \, ,
\end{equation}
where $\bra{n}\hat{H}\ket{n}=\bra{n}\hat{H}_\Lambda\ket{n} \ \forall n$, $\bra{n}\hat{H}\ket{k}=\bra{n}\hat{H}_\Lambda\ket{k}$ if both $n,k\leq \Lambda $, and $\bra{n}\hat{H}_\Lambda\ket{k}=0$ otherwise. We note that by this construction, time evolution under $\hat{H}_\Lambda$ preserves the truncated Hilbert space. We denote the eigenstates and energies of the full Hamiltonian by $\ket{\psi_n}$ and $E_n$, and the eigenstates and energies of the truncated Hamiltonian by $\ket{\psi_n^\Lambda}$ and $E^\Lambda_n$. Thus
\begin{align}
    \label{eq:basic_eqns}
    \ket{\psi_n} &= \ket{\psi_n^\Lambda} + \ket{\delta \psi_n^\Lambda} \nonumber \\
    \ket{\delta \psi_n^\Lambda} & = \Big(1 - \ket{\psi_n^\Lambda}\bra{\psi_n^\Lambda}\Big)\frac{1}{E_n - \hat{H}_\Lambda} \hat{V}_\Lambda \ket{\psi_n} \nonumber \\
    E_n - E_n^\Lambda & = \bra{\psi_n^\Lambda} \hat{V}_\Lambda \ket{\delta \psi_n^\Lambda} \,
\end{align}
where $\bra{\psi_n^\Lambda}\ket{\delta \psi_n^\Lambda} = 0$. Using standard perturbation theory, the leading order correction to $\ket{\psi_n}$ is given by\footnote{Note that $\ket{\psi_n}$ is not normalized to $1$, but when expanding perturbatively, correcting this will only contribute at higher orders in perturbation theory.}
\begin{equation}
    \ket{\delta \psi_n^\Lambda} = \sum_{m > \Lambda} \ket{m} \frac{\bra{m}\hat{V}_\Lambda \ket{\psi_n^\Lambda}}{E^\Lambda_n - \bra{m}\hat{H}_\Lambda \ket{m}} \;.
\end{equation}
Since $\hat{V}_\Lambda$ can only change the boson number by at most 1 and the state $\ket{\psi_n^\Lambda}$ only has support within the truncated Hilbert space, only $m=\Lambda+1$ survives in the sum. Furthermore, since $\ket{\Lambda}$ is the only state within the truncated Hilbert space that is connected to $\ket{\Lambda+1}$ by $\hat{V}_\Lambda$, we can insert $\ket{\Lambda}\bra{\Lambda}$ to write 
\begin{equation}
    \label{eq:correction_term}
    \ket{\delta \psi_n^\Lambda} = \ket{\Lambda+1} \frac{\bra{\Lambda+1}\hat{V}_\Lambda \ket{\Lambda} \bra{\Lambda}\ket{\psi_n^\Lambda}}{E^\Lambda_n - \bra{\Lambda+1}\hat{H}_\Lambda \ket{\Lambda+1}} \, .
\end{equation}
We note that the magnitude of this correction term is controlled by the quantity $\epsilon_{\Lambda}$ defined as follows:
\begin{align}
    \epsilon_{\Lambda} := \frac{\left|\bra{\Lambda+1}\hat{V}_\Lambda \ket{\Lambda}\right|}{\left|E^\Lambda_n - \bra{\Lambda+1}\hat{H}_\Lambda \ket{\Lambda+1}\right|}.
\end{align}
The RHS of \cref{eq:correction_term} depends on the overlap of the truncated eigenstate with the bosonic mode corresponding to the largest bosonic number kept in the truncated Hilbert space. To determine the scaling as the truncation is raised, we can determine this overlap by expanding perturbatively around a lower truncation $\Lambda_0$. First, we note that
\begin{align}
    \bra{\Lambda_0+1}\ket{\psi_n} &= \bra{\Lambda_0+1}\ket{\psi_n^{\Lambda_0}} + \bra{\Lambda_0+1}\ket{\delta \psi_n^{\Lambda_0}} \nonumber \\ &= \bra{\Lambda_0+1}\ket{\delta \psi_n^{\Lambda_0}} \nonumber \\
    &= \frac{\bra{\Lambda_0+1}\hat{V}_{\Lambda_0} \ket{\Lambda_0} \bra{\Lambda_0}\ket{\psi_n^{\Lambda_0}}}{E^{\Lambda_0}_n - \bra{\Lambda_0+1}\hat{H}_{\Lambda_0} \ket{\Lambda_0+1}},
\end{align}
where the second equality follows from the fact that $\psi_n^{\Lambda_0}$ only has support on $\{\ket{0},\ket{1},\dots,\ket{\Lambda_0}\}$ and the third equality follows from \cref{eq:correction_term}. We also have that that
\begin{align}
    \bra{\Lambda_0+1}\ket{\psi_n} = \bra{\Lambda_0+1}\ket{\psi_n^{\Lambda_0+1}} + \bra{\Lambda_0+1}\ket{\delta \psi_n^{\Lambda_0+1}} = \bra{\Lambda_0+1}\ket{\psi_n^{\Lambda_0+1}}
\end{align}
since \cref{eq:correction_term} implies that $\ket{\delta \psi_n^{\Lambda_0+1}} \propto \ket{\Lambda_0+2}$. Putting this together, we thus have that
\begin{align}
    \bra{\Lambda_0+1}\ket{\psi_n} = \bra{\Lambda_0+1}\ket{\psi_n^{\Lambda_0+1}} = \frac{\bra{\Lambda_0+1}\hat{V}_{\Lambda_0} \ket{\Lambda_0} \bra{\Lambda_0}\ket{\psi_n^{\Lambda_0}}}{E^{\Lambda_0}_n - \bra{\Lambda_0+1}\hat{H}_{\Lambda_0} \ket{\Lambda_0+1}}.
\end{align}
Similarly, it is true that
\begin{align}
    \bra{\Lambda_0+2}\ket{\psi_n^{\Lambda_0+2}} &= \frac{\bra{\Lambda_0+2}\hat{V}_{\Lambda_0+1} \ket{\Lambda_0+1} \bra{\Lambda_0+1}\ket{\psi_n^{\Lambda_0+1}}}{E^{\Lambda_0+1}_n - \bra{\Lambda_0+2}\hat{H}_{\Lambda_0+1} \ket{\Lambda_0+2}}  \nonumber \\
    &= \bra{\Lambda_0}\ket{\psi_n^{\Lambda_0}} \cdot \frac{\bra{\Lambda_0+2}\hat{V}_{\Lambda_0+1} \ket{\Lambda_0+1}}{E^{\Lambda_0+1}_n - \bra{\Lambda_0+2}\hat{H}_{\Lambda_0+1} \ket{\Lambda_0+2}} \cdot \frac{\bra{\Lambda_0+1}\hat{V}_{\Lambda_0} \ket{\Lambda_0}}{E^{\Lambda_0}_n - \bra{\Lambda_0+1}\hat{H}_{\Lambda_0} \ket{\Lambda_0+1}} \nonumber \\ 
    &= \bra{\Lambda_0}\ket{\psi_n^{\Lambda_0}} \cdot \frac{\bra{\Lambda_0+2}\hat{V}_{\Lambda_0} \ket{\Lambda_0+1}}{E^{\Lambda_0+1}_n - \bra{\Lambda_0+2}\hat{H}_{\Lambda_0} \ket{\Lambda_0+2}} \cdot \frac{\bra{\Lambda_0+1}\hat{V}_{\Lambda_0} \ket{\Lambda_0}}{E^{\Lambda_0}_n - \bra{\Lambda_0+1}\hat{H}_{\Lambda_0} \ket{\Lambda_0+1}} \nonumber \\
    &\approx \bra{\Lambda_0}\ket{\psi_n^{\Lambda_0}} \cdot \frac{\bra{\Lambda_0+2}\hat{V}_{\Lambda_0} \ket{\Lambda_0+1}}{E^{\Lambda_0}_n - \bra{\Lambda_0+2}\hat{H}_{\Lambda_0} \ket{\Lambda_0+2}} \cdot \frac{\bra{\Lambda_0+1}\hat{V}_{\Lambda_0} \ket{\Lambda_0}}{E^{\Lambda_0}_n - \bra{\Lambda_0+1}\hat{H}_{\Lambda_0} \ket{\Lambda_0+1}}
\end{align}
where we have treated $\epsilon_{\Lambda_0}$ as a small parameter, and the approximate equality above is to leading order in $\epsilon_{\Lambda_0}$ (since \cref{eq:basic_eqns,eq:correction_term} taken together imply that $E^{\Lambda_0+1}_n-E^{\Lambda_0}_n = \mathcal{O}(\epsilon_{\Lambda_0}^2)$). Continuing this process, we can compute $\bra{\Lambda_0+3}\ket{\psi_n^{\Lambda_0+3}}, \bra{\Lambda_0+4}\ket{\psi_n^{\Lambda_0+4}}$, and so on to leading order until we reach the target truncation of $\Lambda$. We thus arrive at the following result:
\begin{align}
    \bra{\Lambda}\ket{\psi_n^\Lambda} = \bra{\Lambda_0}\ket{\psi_n^{\Lambda_0}} \prod_{k=\Lambda_0}^{\Lambda-1} \frac{\bra{k+1}\hat{V}_{\Lambda_0} \ket{k} }{E^{\Lambda_0}_n - \bra{k+1}\hat{H}_{\Lambda_0} \ket{k+1}} \,
\end{align}
to leading order. Therefore, the leading corrections to the states and energies at a truncation of $\Lambda$ are
\begin{align}
    \label{eq:leading_order_corrections}
    \ket{\delta \psi_n^\Lambda} & = \ket{\Lambda+1} \bra{\Lambda_0}\ket{\psi_n^{\Lambda_0}} \frac{\bra{\Lambda+1}\hat{V}_{\Lambda} \ket{\Lambda} }{E^{\Lambda}_n - \bra{\Lambda+1}\hat{H}_{\Lambda} \ket{\Lambda+1}} \prod_{k=\Lambda_0}^{\Lambda-1} \frac{\bra{k+1}\hat{V}_{\Lambda_0} \ket{k} }{E^{\Lambda_0}_n - \bra{k+1}\hat{H}_{\Lambda_0} \ket{k+1}} \\
    E_n - E_n^\Lambda & = \frac{1}{E^\Lambda_n - \bra{\Lambda+1}\hat{H}_{\Lambda_0} \ket{\Lambda+1}} \abs{\bra{\Lambda_0}\ket{\psi_n^{\Lambda_0}} \bra{\Lambda}\hat{V}_{\Lambda_0}\ket{\Lambda+1} \prod_{k=\Lambda_0}^{\Lambda-1} \frac{\bra{k+1}\hat{V}_{\Lambda_0} \ket{k} }{E^{\Lambda_0}_n - \bra{k+1}\hat{H}_{\Lambda_0} \ket{k+1}}}^2 \, .\nonumber 
\end{align}

Treating $\epsilon_{\Lambda_0}$ as a small parameter is valid when $\left|\bra{\Lambda_0+1}\hat{V}_{\Lambda_0} \ket{\Lambda_0}\right| \ll \left|E^{\Lambda_0}_n - \bra{\Lambda_0+1}\hat{H}_{\Lambda_0} \ket{\Lambda_0+1}\right|$. For Kogut-Susskind formulations of lattice gauge theory Hamiltonians we note that the electric energy gap grows quadratically with the size of the representation whereas the plaquette matrix elements are bounded, which makes treating $\epsilon_{\Lambda_0}$ as a small parameter a better and better approximation as we increase the target truncation past $\Lambda_0$ assuming $\Lambda_0$ is chosen to be large enough. For the corrections \cref{eq:leading_order_corrections} to systematically decrease as the truncation is raised, it is necessary for the energy of the eigenstate to be below $\bra{\Lambda_0}\hat{H}_{\Lambda_0} \ket{\Lambda_0}$. For such low energy states, where $E^{\Lambda_0}_n \ll \bra{k+1}\hat{H}_{\Lambda_0} \ket{k+1}$, the size of the corrections will scale as $\mathcal{O}\Big(\prod_{n<\Lambda} \frac{\bra{n+1}\hat{V}_{\Lambda_0}\ket{n}}{\bra{n+1}\hat{H}_{\Lambda_0}\ket{n+1}}\Big)$.

We can make this argument more concrete for the case of $U(1)$ on a single plaquette. We split up the Hamiltonian shown in \cref{eq:EU1def} as follows:
\begin{align}
    \hat{H} & = \hat{H}_\Lambda + \hat{V}_\Lambda\ \,\ ,
\end{align}
with
\begin{align}
    \hat{H}_\Lambda & = 2g^2 \hat{E}^2 + \frac{1}{2g^2} \left(2 - \hat{\Box}_\Lambda - \hat{\Box}_\Lambda^\dagger\right) \nonumber \\
    \hat{\Box}_\Lambda & = \nonumber \sum_{n=-\Lambda+1}^{\Lambda-1} \ket{n} \bra{n+1} \\
    \hat{V}_\Lambda & =  - \frac{1}{2g^2}\sum_{|n|\geq\Lambda} \ket{n} \bra{n+1}  + \text{h.c.}\, .
\end{align}
This implies that for $k \geq \Lambda_0$ we have
\begin{align}
    \label{eq:u1_matrix_elements}
    \langle k+1 | \hat V_{\Lambda_0} | k \rangle &= - \frac{1}{2g^2}\nonumber\\
    \langle k+1 | \hat H_{\Lambda_0} | k+1 \rangle &= 2g^2(k+1)^2\, .
\end{align}
Thus, the product shown in \cref{eq:leading_order_corrections} becomes
\begin{align}
    \prod_{k=\Lambda_0}^{\Lambda-1} \frac{1}{4g^4(k+1)^2} = \mathcal{O}\left(\frac{1}{(\Lambda !)^2}\right).
\end{align}
This expansion is valid for eigenstates where 
\begin{equation}
    \epsilon_{\Lambda_0} = \frac{1}{2g^2 E_n^{\Lambda_0} - 4g^4 (\Lambda_0+1)^2} \ \ \ ,
\end{equation}
is a small parameter.
The continuum limit is the $g\rightarrow0$ limit, which means that $\Lambda_0$ must be taken to be larger for this expansion to be valid as the continuum limit is approached.
For a theory with a boson creation operator and number operator in the Hamiltonian, we have that
\begin{align}
    \langle k+1 | \hat V_{\Lambda_0} | k \rangle &\propto \sqrt{k+1}\nonumber\\
    \langle k+1 | \hat H_{\Lambda_0} | k+1 \rangle &\propto k+1\,,
\end{align}
and hence the the product shown in \cref{eq:leading_order_corrections} becomes
\begin{align}
    \prod_{k=\Lambda_0}^{\Lambda-1} \frac{1}{\sqrt{k+1}} = \mathcal{O}\left(\frac{1}{\sqrt{\Lambda !}}\right).
\end{align}
Note that these expressions were derived for a theory with a single bosonic mode, but it is expected that the general scaling of the truncation corrections should hold for larger systems since the scalings are determined by local properties of the Hamiltonian.

\subsection{Time Dependent Truncation Errors with Fragmented States}

\subsubsection{Single Plaquettes}
\label{sec:OnePlaqAnalytic}

The results of the previous section show that truncation errors in eigenstates go to zero as a factorial of the field truncation, but do not provide quantitative estimates of the truncation error. In this section, it will be shown how time-dependent perturbation theory can be used to estimate truncation errors quantitatively. In the authors' previous work, it was demonstrated that the Kogut--Susskind Hamiltonian generically demonstrates Hilbert space fragmentation (HSF) \cite{Ciavarella:2025zqf}. This is due to the quadratic nature of the electric energy terms, leading to large gaps in the spectrum for states with large electric fields. This will allow us to apply a strong coupling expansion to the dynamics of states with these large electric fields. In the following discussion, a single plaquette will be considered, allowing the entire state of the system to be specified by the electric field on a single link. 
Note that while errors in the time evolution will be estimated here, the techniques in this section can be applied to time-dependent modifications of the Hamiltonian. If this is applied to an adiabatic switching procedure, one can estimate truncation errors in eigenstates~\cite{Gell-Mann:1951ooy}. 
The results of estimating truncation errors in such a manner are consistent with the previous section. Explicitly, to study the evolution of a state $\ket{\phi(t)}$, define the interaction picture state $\ket{\phi_I(t)}$ by
\begin{equation}
    \ket{\phi(t)} = e^{-i\hat{H}_\Lambda t} \ket{\phi_I(t)} \, ,
\end{equation}
where we assume $\ket{\phi(0)}$ only has support on basis states with electric field below $\Lambda$. The state $\ket{\phi_I(t)}$ satisfies
\begin{equation}
    \ket{\phi_I(t)} = \ket{\phi(0)} - i \int_{0}^{t} dt_0 \ e^{i\hat{H}_\Lambda t_0} \hat{V}_\Lambda  e^{-i\hat{H}_\Lambda t_0} \ket{\phi_I(t_0)} \, .
\end{equation}
By repeatedly substituting $\ket{\phi_I(t)}$ into this equation, one can generate a series for $\ket{\phi_I(t)}$ in powers of $\hat{V}_\Lambda$. In particular, we have that
\begin{align}
    \ket{\phi_I(t)} &= \ket{\phi(0)} - i \int_{0}^{t} dt_0 \ e^{i\hat{H}_\Lambda t_0} \hat{V}_\Lambda  e^{-i\hat{H}_\Lambda t_0} \left(\ket{\phi(0)} - i \int_{0}^{t_0} dt_1 \ e^{i\hat{H}_\Lambda t_1} \hat{V}_\Lambda \ket{\phi_I(t_1)} \right) \nonumber \\ &= \ket{\phi(0)} - i \int_{0}^{t} dt_0 \ e^{i\hat{H}_\Lambda t_0} \hat{V}_\Lambda  e^{-i\hat{H}_\Lambda t_0} \ket{\phi(0)} + \mathcal{O}(\hat{V}_{\Lambda}^2)
\end{align}
Thus, 
\begin{align}
    (e^{-i\hat{H}t} - e^{-i\hat{H}_\Lambda t})\ket{\phi(0)} &= e^{-i\hat{H}_\Lambda t}(\ket{\phi_I(t)} - \ket{\phi(0)}) \nonumber \\
    &= - i e^{-i\hat{H}_\Lambda t} \int_{0}^{t} dt_0 \ e^{i\hat{H}_\Lambda t_0} \hat{V}_\Lambda  e^{-i\hat{H}_\Lambda t_0} \ket{\phi(0)} \label{eq:leading_order} \nonumber \\ 
    & = - i e^{-i\hat{H}_\Lambda t} \ket{\Lambda+1} \int_{0}^{t} dt_0 \ e^{i \bra{\Lambda+1}\hat{H}_\Lambda \ket{\Lambda+1} t_0} \bra{\Lambda+1}\hat{V}_\Lambda \ket{\Lambda} \bra{\Lambda}  e^{-i\hat{H}_\Lambda t_0} \ket{\phi(0)} \,
\end{align}
to leading order. Note that similar expressions have been found to govern Trotterization error in low-energy subspaces~\cite{Mizuta:2025aoi}. To determine the size of this expression, the behavior of $\bra{\Lambda}  e^{-i\hat{H}_\Lambda t_0} \ket{\phi(0)}$ needs to be understood. For $\Lambda$ large enough for fragmentation to occur, the phase of this term will be dominated by the diagonal piece of the Hamiltonian so we can approximate $\bra{\Lambda}  e^{-i\hat{H}_\Lambda t_0} \ket{\phi(0)} \approx c_\Lambda \  e^{-i \bra{\Lambda}\hat{H}_\Lambda \ket{\Lambda} t_0}$ where $c_\Lambda$ is a slowly varying function with $\abs{c_\Lambda}^2 \leq 1$. Note that the speed of the evolution of $c_\Lambda = \bra{\Lambda}  e^{-i\hat{H}_\Lambda t_0} \ket{\phi(0)} e^{i \bra{\Lambda}\hat{H}_\Lambda \ket{\Lambda} t_0}$ is set by the energy levels of $\hat{H}_\Lambda$ with the electric energy subtracted off. Due to this subtraction, the fastest phases in $c_\Lambda$ are at most $\mathcal{O}(g^2\Lambda)$ while $ \bra{\Lambda}\hat{H}_\Lambda \ket{\Lambda}=\mathcal{O}(g^2\Lambda^2)$.
With this approximation, we can evaluate this integral and find
\begin{align}
    \left|(e^{-i\hat{H}t} - e^{-i\hat{H}_\Lambda t})\ket{\phi(0)}\right| &\approx |\bra{\Lambda+1}\hat{V}_\Lambda \ket{\Lambda}| \left|\int_0^t dt_0 e^{i (\bra{\Lambda+1}\hat{H}_\Lambda \ket{\Lambda+1} - \bra{\Lambda}\hat{H}_\Lambda \ket{\Lambda}) t_0}\right| \nonumber \\ &= |\bra{\Lambda+1}\hat{V}_\Lambda \ket{\Lambda}| \left|\frac{e^{i (\bra{\Lambda+1}\hat{H}_\Lambda \ket{\Lambda+1} - \bra{\Lambda}\hat{H}_\Lambda \ket{\Lambda}) t}-1}{\bra{\Lambda+1}\hat{H}_\Lambda \ket{\Lambda+1} - \bra{\Lambda}\hat{H}_\Lambda \ket{\Lambda}}\right| \nonumber \\ &\leq \frac{2 \cdot |\bra{\Lambda+1}\hat{V}_\Lambda \ket{\Lambda}|}{\bra{\Lambda+1}\hat{H}_\Lambda \ket{\Lambda+1} - \bra{\Lambda}\hat{H}_\Lambda \ket{\Lambda}}.
\end{align}
The above calculation is valid provided that we truncate at least at the first electric field strength, $\Lambda_0$, where HSF occurs. As in the previous section, the perturbative expansion performed here is valid when $\epsilon_{\Lambda_0}$ is small, where
\begin{equation}
    \epsilon_{\Lambda_0} =  \frac{ |\bra{\Lambda_0+1}\hat{V}_{\Lambda_0} \ket{\Lambda_0}|}{\bra{\Lambda_0+1}\hat{H}_{\Lambda_0} \ket{\Lambda_0+1} - \bra{\Lambda_0}\hat{H}_{\Lambda_0} \ket{\Lambda_0}} \ \ \ .
\end{equation}
However, in practice, we will likely truncate at some $\Lambda > \Lambda_0$ and in this case, we can instead use time-dependent perturbation theory to compute the leading contribution to $\bra{\Lambda}  e^{-i\hat{H}_\Lambda t_0} \ket{\phi(0)}$. Restricting $\ket{\phi(0)}$ to only have support on basis states with electric field below $\Lambda_0$, we start by performing a Dyson series expansion as follows
\begin{align}
    \label{eq:dyson_series}
    e^{-i\hat{H}_\Lambda t_0} \ket{\phi(0)} = e^{-i\hat{H}_{\Lambda_0} t_0} \sum_{N=0}^{\infty}(-i)^N \int_0^{t_0}dt_1\int_0^{t_1}dt_2\dots\int_0^{t_{N-1}}dt_N \hat{V}_{\Lambda_0,I}(t_1) \hat{V}_{\Lambda_0,I}(t_2) \dots \hat{V}_{\Lambda_0,I}(t_N) \ket{\phi(0)},
\end{align}
where we have defined
\begin{align}
    \hat{V}_{\Lambda_0,I}(t_i) := e^{i\hat{H}_{\Lambda_0}t_i}(\hat{H}_{\Lambda} - \hat{H}_{\Lambda_0}) e^{-i\hat{H}_{\Lambda_0}t_i}.
\end{align}
We note that for the truncation levels accessed in these nested integrals, it is true that
\begin{align}
     \hat{V}_{\Lambda_0,I}(t_i) = e^{i\hat{H}_{\Lambda_0}t_i} \hat{V}_{\Lambda_0} e^{-i\hat{H}_{\Lambda_0}t_i} = e^{i\hat{H}_{\Lambda}t_i} \hat{V}_{\Lambda} e^{-i\hat{H}_{\Lambda}t_i}.
\end{align}
Since each application of $(\hat{H}_{\Lambda} - \hat{H}_{\Lambda_0})$ changes the boson number by at most 1, setting $N=\Lambda-\Lambda_0$ results in the leading order contribution to \cref{eq:dyson_series}. Plugging this into the RHS of \cref{eq:leading_order}, we obtain
\begin{align}
    \label{eq:nested_integrals}
    (e^{-i\hat{H}t} - e^{-i\hat{H}_\Lambda t})\ket{\phi(0)} &= -i e^{-i\hat{H}_\Lambda t} \int_0^t dt_0 \  e^{i\hat{H}_\Lambda t_0} \hat{V}_\Lambda e^{-i\hat{H}_\Lambda t_0} (-i)^{\Lambda-\Lambda_0} \nonumber \\ &\times \int_0^{t_0}dt_1\int_0^{t_1}dt_2\dots\int_0^{t_{\Lambda - \Lambda_0-1}}dt_{\Lambda - \Lambda_0} \hat{V}_{\Lambda_0,I}(t_1) \hat{V}_{\Lambda_0,I}(t_2) \dots \hat{V}_{\Lambda_0,I}(t_{\Lambda - \Lambda_0}) \ket{\phi(0)} 
\end{align}
to leading order. This expression consists of integrals of exponentials with phases set by the change in energy from applying $\hat{V}_\Lambda$ or $\hat{V}_{\Lambda_0}$. 
These integrals can be explicitly evaluated as a function of $t$, using the fact that the operator $\hat V_{\Lambda_0}$ only acts on electric basis states $\ket{n}$ with $n \geq \Lambda_0$ and for those states the Hamiltonian $\hat H_{\Lambda_0}$ is diagonal. In particular, the leading order action of $\hat{V}_{\Lambda_0,I}(t_j)$ on a basis state $\ket{\Lambda-j}$ is given by
\begin{align}
    \hat{V}_{\Lambda_0,I}(t_j)\ket{\Lambda-j} &= \bra{\Lambda-j+1}\hat{V}_{\Lambda}\ket{\Lambda-j}e^{i \left(\langle 
    \Lambda - j+1 | \hat{H}_{\Lambda} | \Lambda - j + 1 \rangle - \langle \Lambda-j| H_\Lambda | \Lambda-j \rangle\right)t_j}\ket{\Lambda-j+1} \nonumber \\ &:= \bra{\Lambda-j+1}\hat{V}_{\Lambda}\ket{\Lambda-j}e^{i \left(E_{\Lambda -j + 1}-E_{\Lambda - j}\right)t_j}\ket{\Lambda-j+1},
\end{align}
and thus the norm of the LHS from \cref{eq:nested_integrals} becomes
\begin{align}
    \left|(e^{-i\hat{H}t} - e^{-i\hat{H}_\Lambda t})\ket{\phi(0)}\right|  &= \left|\int_0^t dt_0 e^{i \left(E_{\Lambda + 1}-E_{\Lambda}\right)t_0}\dots\int_0^{t_{\Lambda - \Lambda_0-1}}dt_{\Lambda - \Lambda_0}e^{i \left(E_{\Lambda_0 + 1}-E_{\Lambda_0}\right)t_{\Lambda-\Lambda_0}}\right| \nonumber \\ & \qquad \times \left|\prod_{k=\Lambda_0}^\Lambda \langle k+1 | \hat{V}_{\Lambda} | k \rangle\right| \nonumber\,
    \\ &= \left|\sum_{k=\Lambda_0}^\Lambda  \left(e^{-i \langle 
    \Lambda+1 | \hat{H}_{\Lambda} | \Lambda+1 \rangle t}-e^{-i \langle k| H_\Lambda | k \rangle t}\right)\prod_{\substack{l=\Lambda_0\\l \neq k}}^{\Lambda+1} \frac{1}{ \langle k| H_\Lambda | k \rangle - \langle l| H_\Lambda | l  \rangle}\right| 
    \nonumber\\
    & \qquad \times\left|\prod_{k=\Lambda_0}^\Lambda \langle k+1 | \hat{V}_{\Lambda} | k \rangle\right|  
\,.
\end{align}

These bounds will be independent of what the initial state was, provided that the initial state only has support on basis states with electric fields below $\Lambda_0$. 
In practice, one needs to be able to estimate $\Lambda_0$. A rough way of doing this is to pick $\Lambda_0$ so that the leading order contribution to the leakage error from \cref{eq:leading_order} is smaller than $1$.

To make this discussion concrete, we will apply these results to estimate errors in the truncated simulation of a single plaquette in a $U(1)$ lattice gauge theory. In particular, using \cref{eq:u1_matrix_elements}, we find that
\begin{align}
    \left|(e^{-i\hat{H}t} - e^{-i\hat{H}_\Lambda t})\ket{\phi(0)}\right|  = & \left|\left(\frac{1}{2g^2}\right)^{2(\Lambda+1-\Lambda_0)}
    \sum_{k=\Lambda_0}^\Lambda  \left(e^{-i 2g^2\left(\Lambda+1\right)^2 t}-e^{-i 2g^2k^2 t}\right)\prod_{\substack{l=\Lambda_0\\l \neq k}}^{\Lambda+1} \frac{1}{ k^2-l^2} \right|\nonumber \\
    & \times \left|c_{\Lambda_0} \ket{\Lambda+1}+c_{-\Lambda_0} \ket{-\Lambda-1}\right|\, ,
\end{align}
where $c_{\pm \Lambda_0}$ are the slowly varying norms of $\bra{\pm\Lambda}e^{-i \hat{H_\Lambda}t_{\Lambda_0}}\ket{\phi(0)}$. Defining the leakage amplitude by
\begin{align}
    L(g,\Lambda,\Lambda_0,T) &= \text{max}_{t<T} \abs{ \int_0^t dt_0 e^{i \left(E_{\Lambda + 1}-E_{\Lambda}\right)t_0}\int_0^{t_0}dt_1e^{i \left(E_{\Lambda}-E_{\Lambda-1}\right)t_1}\dots\int_0^{t_{\Lambda - \Lambda_0-1}}dt_{\Lambda - \Lambda_0}e^{i \left(E_{\Lambda_0 + 1}-E_{\Lambda_0}\right)t_{\Lambda-\Lambda_0}}} \, \\ &= \text{max}_{t<T} \abs{ \sum_{k=\Lambda_0}^\Lambda \left[  \left(e^{-i 2g^2\left(\Lambda+1\right)^2 t}-e^{-i 2g^2k^2 t}\right)\prod_{\substack{l=\Lambda_0\\l \neq k}}^{\Lambda+1} \left(\frac{1}{2g^2} \frac{1}{ k^2-l^2} \right) \right]} \, ,
\end{align}
where $T$ is the maximum simulation time, the leading error in the truncated state is thus at most
\begin{equation}
    \abs{(e^{-i\hat{H}t} - e^{-i\hat{H}_\Lambda t})\ket{\phi(0)}} \leq 2 L(g,\Lambda,\Lambda_0,T) \left(\frac{1}{2g^2}\right)^{\Lambda-\Lambda_0+1} \, .
\end{equation}
To get an idea of how this scales, we can place an upper bound on $L(g,\Lambda,\Lambda_0,T)$. In particular, we note that each of the nested integrals from the integral representation of $L(g,\Lambda,\Lambda_0,T)$ can be bounded as
\begin{align}
    \abs{\int_{0}^{t_j-1}dt_je^{i\left(E_{\Lambda-j+1}-E_{\Lambda - j}\right)}} \leq \frac{2}{E_{\Lambda-j+1}-E_{\Lambda - j}} = \frac{1}{g^2(2(\Lambda-j)+1)},
\end{align}
and thus we have that
\begin{align}
    L \leq \prod_{k=\Lambda_0}^{\Lambda}\frac{1}{g^2(2k+1)} = \left(\frac{1}{g^2}\right)^{\Lambda-\Lambda_0+1} \frac{(2\Lambda_0-1)!!}{(2\Lambda+1)!!} \, .
    \label{eq:looseL}
\end{align}
Note that this is a loose bound as it neglects destructive interference between many fast oscillating phases in the actual value of the integral. Regardless, this predicts that the leading error due to truncation converges as a factorial and is time-independent.

\begin{figure}
    \centering
    \includegraphics[width=0.75\linewidth]{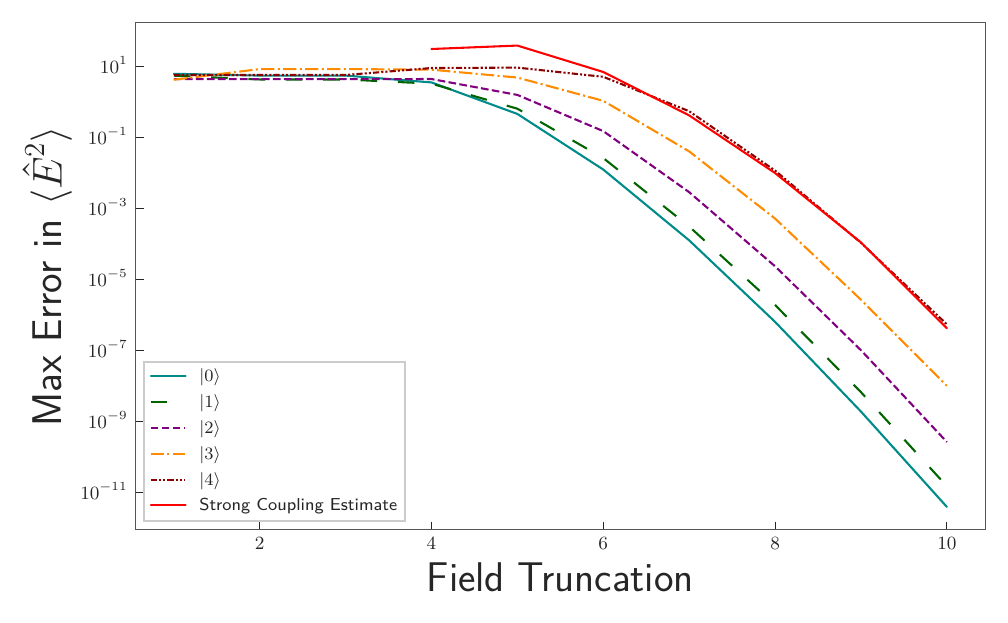}
    \caption{Maximum error in the expectation of $\hat{E}^2$ as a function of the field truncation on a single plaquette with $g=0.5$ and a max evolution time of $T=30$. The blue, green, purple,  orange, and brown lines correspond to using different electric basis states as the initial state. The red curve is the error bound in Eq.~\eqref{eq:OnePlaqE2Bound}, computed using $\Lambda_0 = 4$. }
    \label{fig:OnePlaqEl}
\end{figure}

To understand the performance of these error estimates, we apply these techniques to estimate the error in the expectation of the electric energy $\hat{E}^2$. Using the leading correction to the states, it can be seen that the leading error in the expectation of the electric energy is
\begin{equation}
    \abs{\bra{\phi}e^{i \hat{H}t} \hat{E}^2 e^{-i \hat{H}t}\ket{\phi} -\bra{\phi}e^{i \hat{H}_\Lambda t} \hat{E}^2 e^{-i \hat{H}_\Lambda t}\ket{\phi}} \leq 2 (\Lambda+1)^2 \left(\frac{1}{2g^2}\right)^{\Lambda-\Lambda_0+1} L(g,\Lambda,\Lambda_0,T)^2
    \label{eq:OnePlaqE2Bound}
\end{equation}
We can now compare this bound against the numerical calculation of these expectation values. $\hat{E}^2$ was measured as a function of time for a single plaquette with $g=0.5$ and maximum evolution time of $T=30$. The evolution with $\Lambda=20$ was treated as the exact evolution and was compared to the evolution with lower truncations. The maximum error in the expectation of $\hat{E}^2$ achieved when beginning in different electric basis states, $\ket{n}$, is shown in Fig.~\ref{fig:OnePlaqEl}. As this figure shows, the bound on the error from Eq.~\eqref{eq:OnePlaqE2Bound} correctly upper bounds the actual error for all of these initial states. Note that the bound becomes tighter for initial states with larger electric energies.

\FloatBarrier
\subsubsection{Comparison to Previous Work}
Previous work has derived resource estimates for bosonic theories through a combination of rigorous upper bounds on the Dyson series expansion in an interaction picture and constraints from energy conservation~\cite{Tong:2021rfv}.
To understand how the current work compares to these results, we will study the analytic bounds of this previous work for the evolution of the electric vacuum state on a single plaquette with a $U(1)$ gauge field. Note that these techniques apply to larger system sizes, but for comparison, we will restrict to a single plaquette.
The previous bounds on truncation errors were derived by placing upper bounds on the expectation of $\hat{\Pi}_n = \ket{n} \bra{n}$ as a function of time. These leakage bounds were then turned into a bound on error in the time evolution operator due to truncation. 

One starts from the expectation value of the full Hamiltonian in the electric vacuum $\ket{0}$, which by energy conservation has to be time invariant.
Working again with a U(1) pure gauge theory, the total energy is always greater than the electric energy; one therefore obtains
\begin{align}
    \bra{0}\hat{H} \ket{0} = \bra{0}e^{i \hat{H}t}\hat{H} e^{-i \hat{H}t}\ket{0} = \frac{1}{g^2}
    \, .
\end{align}
Since both the electric and magnetic terms in the Hamiltonian are positive definite, we can bound the total energy from above by the electric energy
\begin{align}
\bra{0}e^{i \hat{H}t}\hat{H} e^{-i \hat{H}t}\ket{0} \geq \bra{0}e^{i \hat{H}t}\hat{H}_E e^{-i \hat{H}t}\ket{0} = 2g^2 \bra{0}e^{i \hat{H}t} \hat{E}^2 e^{-i \hat{H}t} \ket{0}
    \, ,
\end{align}
Using the expression of the electric operator in~\cref{eq:EU1def}, we can bound the electric energy from above by the contributions of a single bosonic state $\ket{\Lambda}$ such giving
\begin{align}
    \bra{0}e^{i \hat{H}t} \hat{E}^2 e^{-i \hat{H}t} \ket{0} \geq \Lambda^2 \bra{0}e^{i \hat{H}t} \hat{\Pi}_\Lambda e^{-i \hat{H}t} \ket{0}\, .
\end{align}
Combining everything together, we obtain the bound
\begin{align}
    \frac{1}{g^2} \geq 2 g^2\Lambda^2 \bra{0}e^{i \hat{H}t} \hat{\Pi}_\Lambda e^{-i \hat{H}t} \ket{0}
    \, ,
\end{align}
or 
\begin{align}
    \label{eq:energy_conservation_prob_bound}
    \bra{0} e^{i \hat{H}t} \hat{\Pi}_\Lambda e^{-i \hat{H}t} \ket{0} & \leq \frac{1}{2g^4 \Lambda^2} \, ,
\end{align}
for all times $t$. 
Note that for a larger lattice, similar volume-independent bounds can be derived for translationally invariant states. 

For short times, a time-dependent bound can be derived by upper-bounding the error in a Dyson series expansion. Explicitly, one works in the interaction picture where the free part of the Hamiltonian is given by $2g^2 \hat{E}^2$. The Hamiltonian in the interaction picture is then given by
\begin{equation}
    \hat{H}_I(t) = - \frac{1}{2g^2}\sum_n e^{i2g^2 t (2n+1)} \ket{n+1}\bra{n} + \text{h.c.} \, .
\end{equation}
Denoting the electric vacuum evolved in the interaction picture by $\ket{\phi_I(t)}$, the expectation of the projector $\hat{\Pi}_\Lambda$ is given by
\begin{align}
    \bra{0} e^{i\hat{H}t}\hat{\Pi}_\Lambda e^{-i\hat{H}t} \ket{0} = \abs{\bra{\Lambda} e^{-i\hat{H}t} \ket{0}}^2 = \abs{\bra{\Lambda} \ket{\phi_I(t)}}^2 \, .
\end{align}
The Dyson series expansion is generated by inserting $\ket{\phi_I(t)}$ recursively into the equation
\begin{equation}
    \ket{\phi_I(T)} = \ket{\phi_I(0)} - i \int_0^T dt \ \hat{H}_I(t) \ket{\phi_I(t)} \, .
\end{equation}
The quantity $\bra{\Lambda} \ket{\phi_I(t)}$ only receives a non-zero contribution from the $\Lambda$-th order which is given by
\begin{equation}
    \bra{\Lambda} \ket{\phi_I(t)} = (-i)^\Lambda \int_0^t dt_1 \int_0^{t_1} dt_2 \cdots \int_0^{t_{\Lambda-1}} dt_{\Lambda}  \hat{H}_I(t_1) \hat{H}_I(t_2) \cdots \hat{H}_I(t_\Lambda)  \ket{\phi_I(t_\Lambda)} \, .
\end{equation}
Since the spectral norm of $H_I(t)$ is bounded by 
\begin{align}
\left|\left|H_I(t)\right|\right| \leq \frac{1}{g^2}
    \, ,
\end{align}
the magnitude of this overlap is upper-bounded by
\begin{align}
    \abs{\bra{\Lambda} \ket{\phi_I(t)}} & \leq \left|\left|H_I(t)\right|\right|^\Lambda \int_0^t dt_1 \int_0^{t_1} dt_2 \cdots \int_0^{t_{\Lambda-1}} dt_{\Lambda} = \frac{t^\Lambda}{g^{2\Lambda} \Lambda!} 
    \, .
\end{align}
We therefore find
\begin{align}
      \bra{0} e^{i\hat{H}t}\hat{\Pi}_\Lambda e^{-i\hat{H}t} \ket{0} & \leq \frac{t^{2\Lambda}}{g^{4\Lambda} \Lambda!^2} 
      \, .  
\end{align}
Similar techniques can be used to show
\begin{equation}
    \abs{\hat{\Pi}_{\Lambda} e^{-i\hat{H}t}\hat{\Pi}_{\Lambda-\Delta}} \leq \frac{t^\Delta}{g^{2\Delta} \Delta!} \, ,
\end{equation}
 where $\Delta$ is the change in the electric field. These results are referred to as short-time bounds, as at long times they will exceed the bound derived from energy conservation. 

Time-dependent bounds that are valid for longer times were also derived by inserting intermediate projectors and bounding individual terms in the sum. 
For example, one can consider adding one intermediate projector, defining
\begin{align}
    \hat{\Pi}_{>\Lambda} & = \sum_{\abs{n}>\Lambda} \ket{n}\bra{n} \nonumber \\
    \hat{\Pi}_{\leq\Lambda} & = \sum_{\abs{n}\leq\Lambda} \ket{n}\bra{n} \, .
\end{align}
This allows to derive
\begin{align}
    \abs{\hat{\Pi}_{\Lambda} e^{-i\hat{H}t}\hat{\Pi}_{\Lambda-\Delta_1-\Delta_2}} & = \abs{\hat{\Pi}_{\Lambda} e^{-i\hat{H}(t-t_1)} \left(\hat{\Pi}_{>\Lambda-\Delta_1} + \hat{\Pi}_{\leq\Lambda-\Delta_1}\right) e^{-i\hat{H}t_1}\hat{\Pi}_{\Lambda-\Delta_1-\Delta_2}} \nonumber \\
    & \leq \abs{\hat{\Pi}_{>\Lambda-\Delta_1}e^{-i\hat{H}t_1}\hat{\Pi}_{\Lambda-\Delta_1-\Delta_2}} + \abs{\hat{\Pi}_{\Lambda} e^{-i\hat{H}(t-t_1)}\hat{\Pi}_{\leq\Lambda-\Delta_1}} \nonumber \\
    & \leq \frac{t_1^{\Delta_2}}{g^{2{\Delta_2}} {\Delta_2}!} + \frac{(t-t_1)^{\Delta_1}}{g^{2{\Delta_1}} {\Delta_1}!} \, ,
\end{align}
and $t_1$ can be chosen to minimize this expression. 
To obtain a long-term bound for the expectation value $\bra{0} e^{i\hat{H}t}\hat{\Pi}_\Lambda e^{-i\hat{H}t} \ket{0}$ one can insert up to $\Lambda$ projectors and optimize over the intermediate times. 
These are referred to as long-time bounds, as they take longer to saturate the bound from energy conservation.

\begin{figure}
    \centering
    \includegraphics[width=\linewidth]{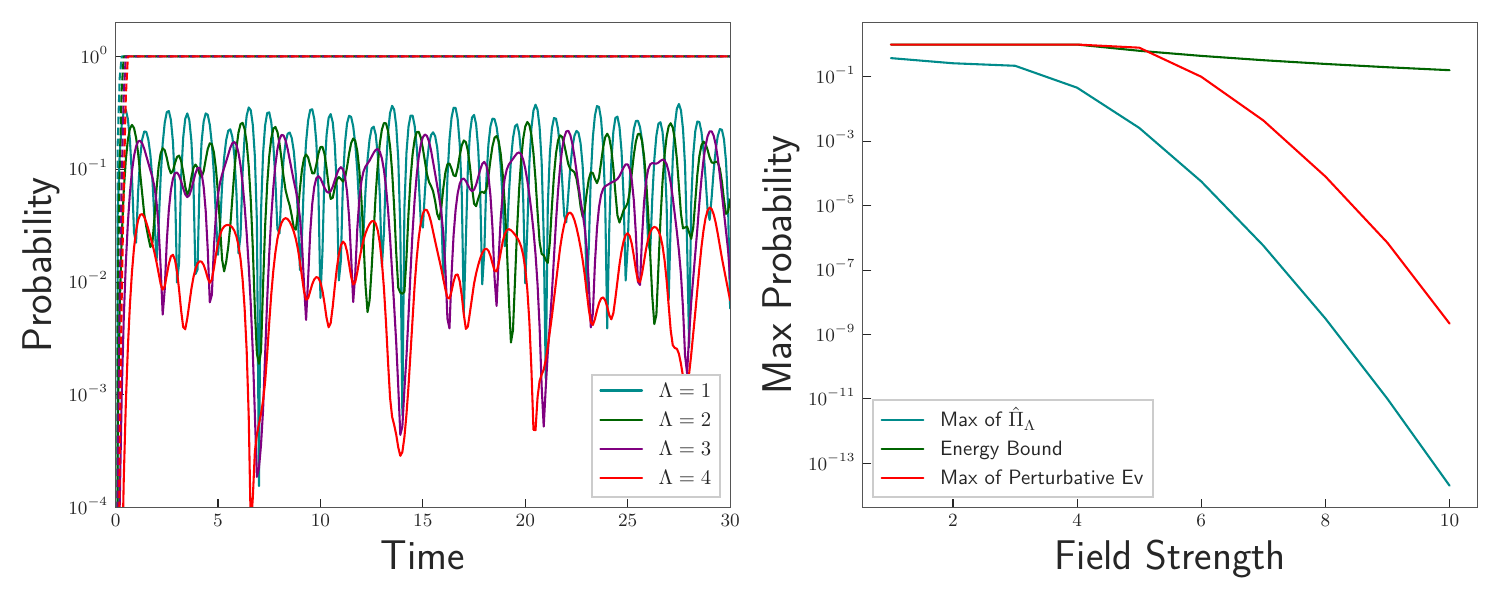}
    \caption{Evolution of the electric vacuum state on a single plaquette. Numerical simulations were performed with a maximum electric field of $20$ and $g=0.5$. The left panel shows the expectation of $\hat{\Pi}_\Lambda$ for various values of $\Lambda$ as a function of time $t$. The solid curves are the exact time evolution, and the dashed lines are the rigorous long-time bounds. The right panel shows the maximum of the expectation of $\hat{\Pi}_\Lambda$ for the simulated time evolution. The blue curve is the exact result, the green curve is the bound from energy conservation, and the red curve is the leading contribution to the expectation obtained by calculating $L(g,\Lambda,\Lambda_0,T)$ with $\Lambda_0=4$.}
    \label{fig:OnePlaq}
\end{figure}

In summary, the work of~\cite{Tong:2021rfv} derived three types of bounds, namely time-independent bounds based on energy conservation, as well as time-dependent bounds valid at short and longer times. 
These can be combined into a more optimal time-dependent bound by taking their minimum.
To determine the tightness of the resulting bound, the time evolution of the electric vacuum state will be simulated numerically. Note that 
\begin{equation}
    \bra{0}e^{i\hat{H}t} \hat{\Pi}_\Lambda e^{-i\hat{H}t} \ket{0} \leq \abs{\hat{\Pi}_\Lambda e^{-i\hat{H}t} \hat{\Pi}_0}^2 \, , 
\end{equation}
so these bounds directly translate to bounds on the expectation of operators.
The electric vacuum was evolved under a Hamiltonian with a truncation of $\Lambda=20$ and $g=0.5$. The solid lines in the left panel of Fig.~\ref{fig:OnePlaq} show the exact time evolution of the expectation of $\hat{\Pi}_\Lambda$ and the dashed lines show the bounds as derived above (the minimum of the three approaches discussed). As this figure shows, the long-time bounds quickly saturate and overestimate the expectation of $\hat{\Pi}_\Lambda$.
The figure on the right shows the behavior of the maximum of the expectation value of the projection operator taken over large times, as a function of the truncation $\Lambda$.  
In blue, we show the result from the numerical simulation, which shows that the maximum expectation for $\Lambda = 10$ is below $10^{-13}$.
The green curve is the result of ~\cite{Tong:2021rfv} derived in this section, which clearly overestimates the true value by many orders of magnitude. 
We also show in red the new results of this work, which were derived in the previous section. 
One can see that these results provide a much tighter bound.

\begin{figure}
    \centering
    \includegraphics[width=0.75\linewidth]{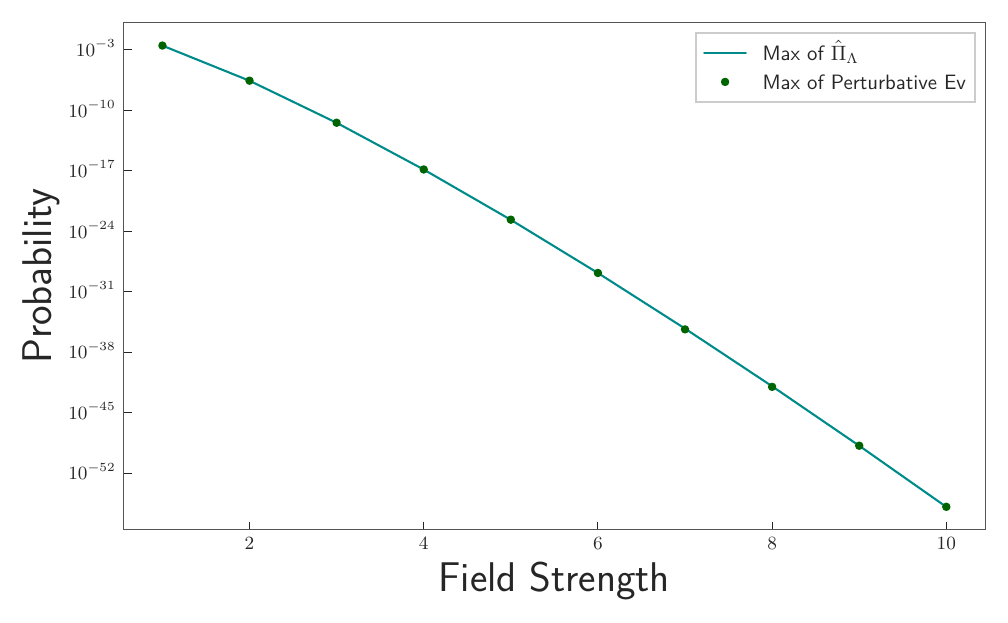}
    \caption{Evolution of the electric vacuum state on a single plaquette. Numerical simulations were performed with a maximum electric field of $100$, $g=\sqrt{3}$, and a maximum evolution time of $t=8$. The blue curve is the maximum of the expectation of $\hat{\Pi}_\Lambda$ during this evolution. The green points are the leading contribution to the expectation obtained by calculating $L(g,\Lambda,\Lambda_0,T)$ with $\Lambda_0=0$.}
    \label{fig:OnePlaqComparison}
\end{figure}

As discussed in~\cite{Tong:2021rfv}, the bounds derived in that work imply that  to guarantee $\abs{\bra{k+\Lambda}e^{-i\hat{H}t}\ket{k}}<0.01$ for $g=\sqrt{3}$ for a maximum evolution time of $t=8$ would require a cutoff satisfying $\Lambda>100$. 
On the other hand, using the upper bound derived in this work for $L(g,\Lambda,\Lambda_0,T)$, given in   Eq.~\eqref{eq:looseL}, one estimates that $\abs{\bra{k+\Lambda}e^{-i\hat{H}t}\ket{k}}<6\times10^{-308}$ for these parameters.
This shows in a pretty dramatic fashion how much tighter the bounds derived in this work are. 
To verify this claim numerically, the evolution of the electric vacuum on a single plaquette was simulated for $g=\sqrt{3}$ and a maximum evolution time of $t=8$. Fig.~\ref{fig:OnePlaqComparison} shows the maximum value reached by the expectation of $\hat{\Pi}_\Lambda$ during this time evolution and the predicted maximum for the perturbative calculation done in this work. As this figure shows, the predicted maximum is in good agreement with the numerical simulation, indicating the tightness of the bounds derived in this work.

The bounds obtained in Ref~\cite{Tong:2021rfv} are extremely loose due to the bounds being computed by applying the triangle inequality to upper-bound the Dyson series expansion. 
The Dyson series consists of a sum over many oscillating phases, and the actual physics of the system comes from how these phases constructively or destructively interfere. 
As the triangle inequality removes these phases, it should not be surprising that these bounds fail to come remotely close to the qualitative behavior of the system. 
The estimate for the maximum of $\hat{\Pi}_\Lambda$ obtained using the strong coupling expansion appears to fall off at the same rate as the actual system, but is still a loose overestimate. This is because the strong coupling expansion cannot be applied to states with small electric fields, and the amplitude for leakage into the fragmented sector of the theory was upper-bounded by $1$.

\subsection{Extension to Larger Systems}
To extract new predictions from simulations of lattice gauge theories on quantum computers, it is necessary to simulate lattices with more than one plaquette and have estimates of the errors due to truncations of the gauge fields. 
This can be done using similar techniques to the single plaquette case. We will write the Hamiltonian for the untruncated theory, $\hat{H}$, as a sum of the truncated Hamiltonian, $\hat{H}_\Lambda$, plus a sum over $\hat{V}^{\Vec{x}}_\Lambda$ which contains all off-diagonal couplings for electric fields above the truncation $\Lambda$ for the plaquette at position $\Vec{x}$, explicitly
\begin{equation}
    \hat{H} = \hat{H}_\Lambda + \sum_x \hat{V}_{\Lambda}^{\Vec{x}} \, .
\end{equation}
Following the same approach as Section~\ref{sec:OnePlaqAnalytic}, the leading correction to the truncated evolution is given by
\begin{equation}
        (e^{-i\hat{H}t} - e^{-i\hat{H}_\Lambda t})\ket{\phi(0)} = \sum_{\Vec{x}} - i e^{-i\hat{H}_\Lambda t} \int_{0}^{t} ds \ e^{i\hat{H}_\Lambda s} \hat{V}^{\Vec{x}}_\Lambda  e^{-i\hat{H}_\Lambda s} \ket{\phi(0)}
        \, .
\end{equation}
The quantity $\hat{V}^{\Vec{x}}_\Lambda  e^{-i\hat{H}_\Lambda s} \ket{\phi(0)}$ can be estimated to leading order by applying time-dependent perturbation theory in the fragmented subspace as before to give at leading order
\begin{align}
\label{eq:evolve_error_large_system}
    & (e^{-i\hat{H}t} - e^{-i\hat{H}_\Lambda t})\ket{\phi(0)}  = -i e^{-i\hat{H}_\Lambda t} \sum_{\Vec{x}} \int_0^t dt_0 \  e^{i\hat{H}_\Lambda t_0} \hat{V}^{\Vec{x}}_\Lambda e^{-i\hat{H}_\Lambda t_0} (-i)^{\Lambda-\Lambda_0} \int_{0}^{t_0} dt_1 e^{i\hat{H}_{\Lambda_0} t_1} \hat{V}^{\Vec{x}}_{\Lambda_0}  e^{-i\hat{H}_{\Lambda_0} t_1} \nonumber \\
    & \times \int_{0}^{t_1} dt_2 e^{i\hat{H}_{\Lambda_0} t_2} \hat{V}^{\Vec{x}}_{\Lambda_0}  e^{-i\hat{H}_{\Lambda_0} t_2} \cdots \int_{0}^{t_{\Lambda-\Lambda_0+1}} dt_{\Lambda-\Lambda_0} e^{i\hat{H}_{\Lambda_0} t_{\Lambda-\Lambda_0}} \hat{V}^{\Vec{x}}_{\Lambda_0}  e^{-i\hat{H}_{\Lambda_0} t_{\Lambda-\Lambda_0}} \ket{\phi(0)} 
    \, .
\end{align}
As before, the integrand in this expression consists of phases given by changes in energy from applying $\hat{V}^{\Vec{x}}_{\Lambda_0}$ or $\hat{V}^{\Vec{x}}_{\Lambda}$. These changes in energy can be approximated by the change in electric energy. 
This gives the leading corrections to the state vector and can be used to evaluate the leading error in the expectation of observables. 
Note that at leading order, local observables will only have errors coming from the $\hat{V}^{\Vec{x}}_{\Lambda}$ terms that share support with the observable.

To make this discussion concrete, the truncation error on the expectation value of an operator will be estimated for a pure $U(1)$ lattice gauge theory on a plaquette chain. 
The Hamiltonian is given by
\begin{equation}
    \hat{H} = \sum_l \frac{g^2}{2} \hat{E}^2_l + \sum_p \frac{1}{2g^2} \left(2 - \hat{\Box}_p - \hat{\Box}^\dagger_p\right) 
    \, ,
\end{equation}
where the sum over $l$ corresponds to summing over all links on the lattice and the sum over $p$ corresponds to summing over all plaquettes on the lattice. 
This Hamiltonian can be gauge-fixed so gauge-invariant states are given by specifying the electric field on a single plaquette per link~\cite{Ciavarella:2025zqf}. The gauge-fixed Hamiltonian is given by
\begin{align}
    \hat{H} &= \sum_p \bigg\{g^2 \left[ \hat{E}_p^2 + \frac{1}{2}\Big(\hat{E}_p - \hat{E}_{p+1}\Big)^2\right]+ \frac{1}{2g^2} \left(2-\hat{\Box}_p - \hat{\Box}^\dagger_p \right)\bigg\} \nonumber \\
    \hat{E}_p &= \sum_{n_p=-\infty}^{\infty} n_p \ket{n_p} \bra{n_p} \nonumber \\
    \hat{\Box}_p &= \sum_{n_p=-\infty}^{\infty} \ket{n_p} \bra{n_p +1  } \, ,
\end{align}
and for a chain of length $L$ the electric basis states are given by $\ket{n_1,n_2,\cdots n_L}$. 
The operator we will choose for this example will be the electric energy on the four links of a given plaquette, and for concreteness, we choose the time-dependent expectation value of $\hat{E}_p^2$. 
This operator will receive contributions from plaquette $p$, as well as the neighboring plaquettes $p\pm1$. 
This implies that the truncation errors can be estimated using a $3$ plaquette chain, with a generic state given by $\ket{n_1, n_2, n_3}$. 
To simplify the notation in the calculation, only the error from the truncation on positive electric fields will be estimated (the contribution from negative electric fields will only contribute a factor of $2$). 

To start, we write as before
\begin{align}
    \ket{\phi(0)} = \sum_{n_1,n_2,n_3 \leq \Lambda_0} c_{n_1,n_2,n_3}\ket{n_1, n_2, n_3}
    \,.
\end{align}
Following similar steps as before, the time-evolved perturbation can be written as
\begin{equation}
    e^{i\hat{H}_{\Lambda_0} t} \hat{V}^{(2)}_{\Lambda_0}  e^{-i\hat{H}_{\Lambda_0} t} \ket{\phi(0)} = \frac{-1}{2g^2}\sum_{n_1,n_3} c_{n_1,\Lambda_0,n_3}e^{i t g^2 (4 \Lambda_0 -n_1 - n_3 + 2)} \ket{n_1,\Lambda_0+1,n_3} 
    \,.
\end{equation}

Defining the function
    \begin{equation}
    A_{n, m}(g,\Lambda,\Lambda_0,T) = \left(\frac{-i}{2g^2}\right)^{\Lambda-\Lambda_0+1} \prod_{k=\Lambda_0}^{\Lambda} \int_{0}^{t_{k+1}} \!\! {\rm d}t_k \, e^{i g^2 (4 k -n - m + 2) t_k} 
    \, ,
    \label{eq:Anm}
\end{equation}

with $T \equiv t_{\Lambda+1}$, the correction to the state vector is therefore given by
\begin{equation}
    (e^{-i\hat{H}t} - e^{-i\hat{H}_\Lambda t})\ket{\phi(0)}  = \sum_{n,m} A_{n,m}(g,\Lambda,\Lambda_0,t) c_{n,\Lambda_0,m} \ket{n,\Lambda+1,m} 
    \,.
\end{equation}
Combining this, we obtain the correction to the expectation of the electric energy on the center plaquette ($\hat{E}^2_2$),
\begin{align}
    &\bra{\phi}e^{i \hat{H}t} \hat{E}_2^2 e^{-i \hat{H}t}\ket{\phi} -\bra{\phi}e^{i \hat{H}_\Lambda t} \hat{E}_2^2 e^{-i \hat{H}_\Lambda t}\ket{\phi} \nonumber\\
    & \qquad\qquad\qquad = \sum_{n,m \leq \Lambda_0} \abs{A_{n,m}(g,\Lambda,\Lambda_0,t)}^2\left[(\Lambda+1)^2  \abs{c_{n,\Lambda_0,m}}^2 +  m^2 \left(\abs{c_{\Lambda_0,m,n} }^2 + \abs{c_{n,m,\Lambda_0} }^2 \right) \right]
    \,.
\end{align}
Using the fact that $\sum_{n_1,n_2,n_3 \leq \Lambda_0} \abs{c_{n_{1},n_{2},n_3}}^2 = 1$, the error can be upper bounded by
\begin{align}
    \abs{\bra{\phi}e^{i \hat{H}t} \hat{E}_2^2 e^{-i \hat{H}t}\ket{\phi} -\bra{\phi}e^{i \hat{H}_\Lambda t} \hat{E}_2^2 e^{-i \hat{H}_\Lambda t}\ket{\phi}}  \leq &\text{max}_{\tau < t} \left\{(\Lambda+1)^2 \abs{A_{n,m}(g,\Lambda,\Lambda_0,\tau)}^2 : n,m \leq \Lambda_0 \right\} \nonumber \\
    & + 2 \text{max}_{\tau < t} \left\{n^2 \abs{A_{n,m}(g,\Lambda,\Lambda_0,\tau)}^2 : n,m \leq \Lambda_0 \right\} 
    \, .
\end{align}
The above expression is only for truncating the positive electric fields with strength above $\Lambda$. When truncating the positive and negative electric fields with strength above $\Lambda$, the resulting expression for the leading error in the electric field evolution is given by
\begin{align}
        \abs{\bra{\phi}e^{i \hat{H}t} \hat{E}_2^2 e^{-i \hat{H}t}\ket{\phi} -\bra{\phi}e^{i \hat{H}_\Lambda t} \hat{E}_2^2 e^{-i \hat{H}_\Lambda t}\ket{\phi}}  \leq & 2\text{max}_{\tau < t} \left\{(\Lambda+1)^2 \abs{A_{n,m}(g,\Lambda,\Lambda_0,\tau)}^2 : \abs{n},\abs{m} \leq \Lambda_0 \right\} \nonumber \\
    & + 4 \text{max}_{\tau < t} \left\{n^2 \abs{A_{n,m}(g,\Lambda,\Lambda_0,\tau)}^2 : \abs{n},\abs{m} \leq \Lambda_0 \right\} \, .
    \label{eq:PlaqChainElError}
\end{align}
Note that while the expression was derived using a three-plaquette lattice, the result will generalize to a plaquette chain of arbitrary length. The only information used about the initial state was that it did not have support on electric fields above $\Lambda_0$. This bound could potentially be made tighter by using energy conservation arguments to upper bound the probability of having a state with an electric field of $\Lambda_0$ present. For translationally invariant states, these energy bounds will have no volume dependence.

This bound is based on a perturbative expansion of the dynamics connecting different fragmented sectors of the Hilbert space. This is an expansion in powers of
\begin{equation}
    \epsilon_{\Lambda_0} = \frac{1}{4g^4(\Lambda_0+1)} \ \ \ .
\end{equation}
Provided that $\Lambda_0$ is taken to be sufficiently large, this bound can be used to estimate truncation errors at any value of $g$, including when $g$ is small enough to probe continuum physics. Note that this means as $g\rightarrow0$, one must have $\Lambda$ grow as $g^{-4}$.

\begin{figure}
    \centering
    \includegraphics[width=\linewidth]{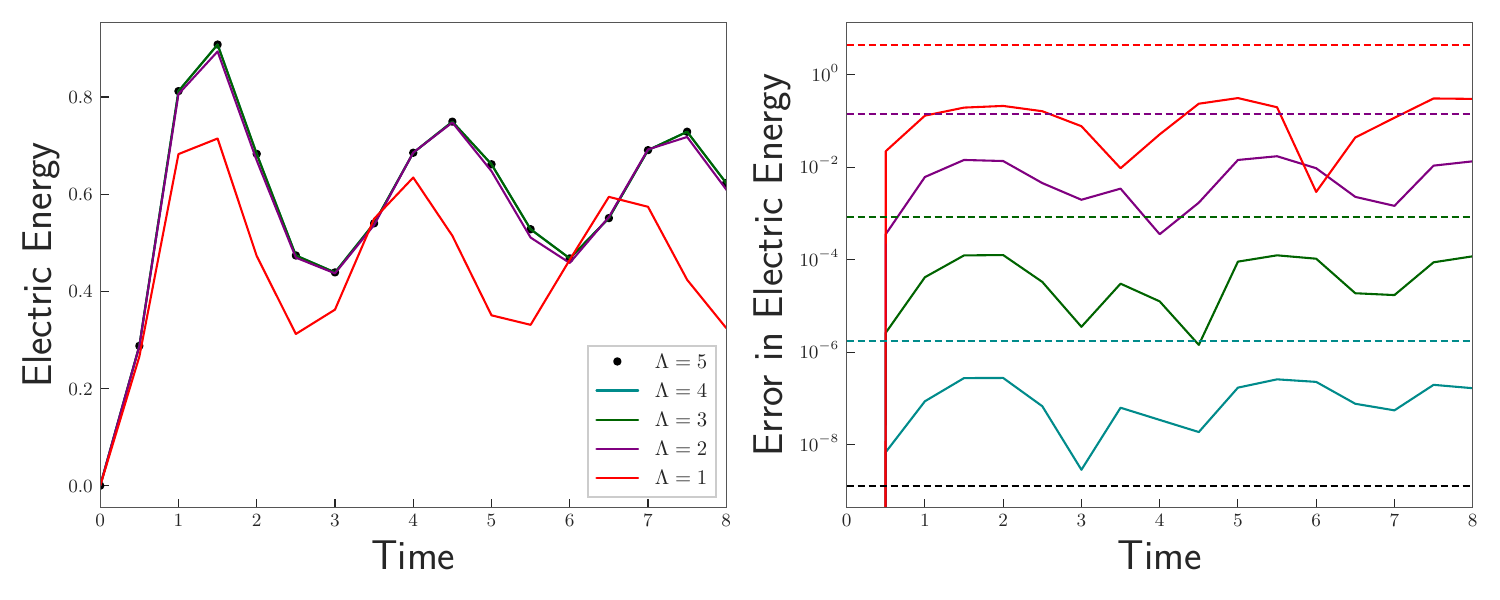}
    \caption{Simulation of an infinite plaquette chain with $g=0.8$. The left panel shows the electric energy as a function of time for different truncations of the electric field. The solid curves in the right panel show the difference in electric energy between the $\Lambda=5$ truncation and the lower truncations. The dashed lines show the predicted error in the electric energy from truncating the Hamiltonian, computed using Eq.~\eqref{eq:PlaqChainElError} with $\Lambda_0=1$. See \cref{app:iMPS} for details regarding the iMPS implementation.}
    \label{fig:plaq-infinite}
\end{figure}

To demonstrate the performance of these error estimates, an infinite MPS was used to simulate the time evolution of an infinite plaquette chain with $g=0.8$~\cite{Vidal:2003lvx}. The system was initialized in the electric vacuum and evolved until $t=8$. The details of the implementation are in Appendix~\ref{app:iMPS}. Fig.~\ref{fig:plaq-infinite} shows the expectation of $\hat{E}^2$ for a single plaquette on the lattice as a function of time. The truncation error in the time evolution is consistent with the estimate of the error computed in Eq.~\eqref{eq:PlaqChainElError}. This agreement suggests that truncation errors in local observables do not have a system size dependence as predicted from the perturbative calculation.

\begin{figure}
    \centering
    \includegraphics[width=\linewidth]{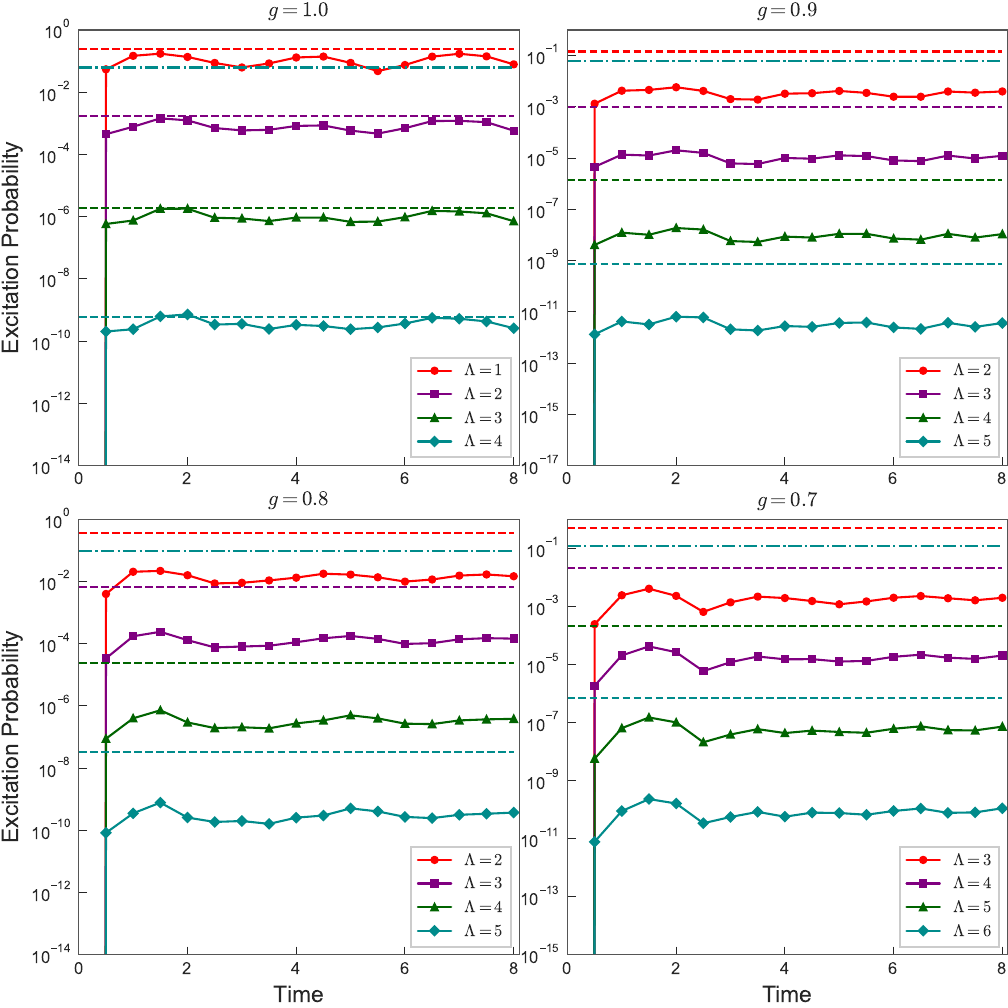}
    \caption{Simulation of an infinite plaquette chain with four different values of $g = 1.0, 0.9, 0.8$, and $0.7$ and maximum electric fields of $5,6,6,$ and $7$ respectively. The solid curves in each panel show the expectation value of the projection operator $\hat{{\Pi}}_{\Lambda,\vec{x}}$ with respect to the states obtained by time evolving the electric vacuum for various choices of $\Lambda$.  See \cref{app:iMPS} for details regarding the iMPS time evolution. The dashed lines in each panel correspond to the upper bound on the probability given by \cref{eq:prob_upper_bound} with $\Lambda_0 = 0,1, 1$, and $2$ respectively for each choice of $g$. The dash-dotted lines in each panel show the upper bound based on energy conservation, computed using \cref{eq:energy_conservation_prob_bound} for the largest value of $\Lambda$ considered for each value of g.}
    \label{fig:u1_prob}
\end{figure}

The probability of exciting a given link on the lattice to a large electric field on a plaquette chain can be estimated using this framework. Denoting $\hat{\Pi}_{\Lambda,\Vec{x}} = \ket{\Lambda_{\Vec{x}}} \bra{\Lambda_{\Vec{x}}}$, the probability of exciting a link to electric field $\Lambda$ when the initial state, $\ket{\phi}$ has all electric fields below $\Lambda_0$ is upper bounded by
\begin{equation}
    \bra{\phi} e^{i \hat{H} t} \hat{\Pi}_{\Lambda,\Vec{x}} e^{-i \hat{H} t} \ket{\phi}\leq P(\Lambda,\Lambda_0,g,t) =\text{max}_{\tau < t} \left\{\abs{A_{n,m}(g,\Lambda,\Lambda_0,\tau)}^2 : \abs{n},\abs{m} \leq \Lambda_0 \right\}
\end{equation}
provided that HSF is present for states with electric fields above $\Lambda_0$. From Eq.~\eqref{eq:Anm}, it follows that $P(\Lambda,\Lambda_0,g,t)$ is upper bounded by
\begin{equation}
    \label{eq:prob_upper_bound}
    P(\Lambda,\Lambda_0,g,t) \leq \frac{1}{g^{8(\Lambda-\Lambda_0)}} \times f(\Lambda,\Lambda_0) \,,
\end{equation}
where $f(\Lambda,\Lambda_0)$ is a $g$ independent function. \cref{fig:u1_prob} shows a numerical demonstration of the validity of this upper bound. Using an infinite MPS, we again simulate the time evolution of an infinite plaquette chain with four different values of $g = 1.0, 0.9, 0.8$, and $0.7$. Starting from the electric vacuum, we perform time evolution until $t=8$ and compute the expectation value $\langle \hat{\Pi}_{\Lambda,\vec{x}} \rangle$ as a function of time. We observe that our numerical results are consistent with \cref{eq:prob_upper_bound}. For further implementation details pertaining to the time evolution, see \cref{app:iMPS}. \cref{fig:u1_prob} also shows a comparison between the bound shown in \cref{eq:prob_upper_bound} and the time independent energy conservation based bound shown in \cref{eq:energy_conservation_prob_bound}. In particular, we show that for the largest value of $\Lambda$ considered for each value of $g$, the bound based on energy conservation is between $5$ and $8$ orders of magnitude larger than the bound based on \cref{eq:prob_upper_bound}.

\FloatBarrier
\section{Truncated Schwinger Model}
\begin{figure}
    \centering
    \includegraphics[width=\linewidth]{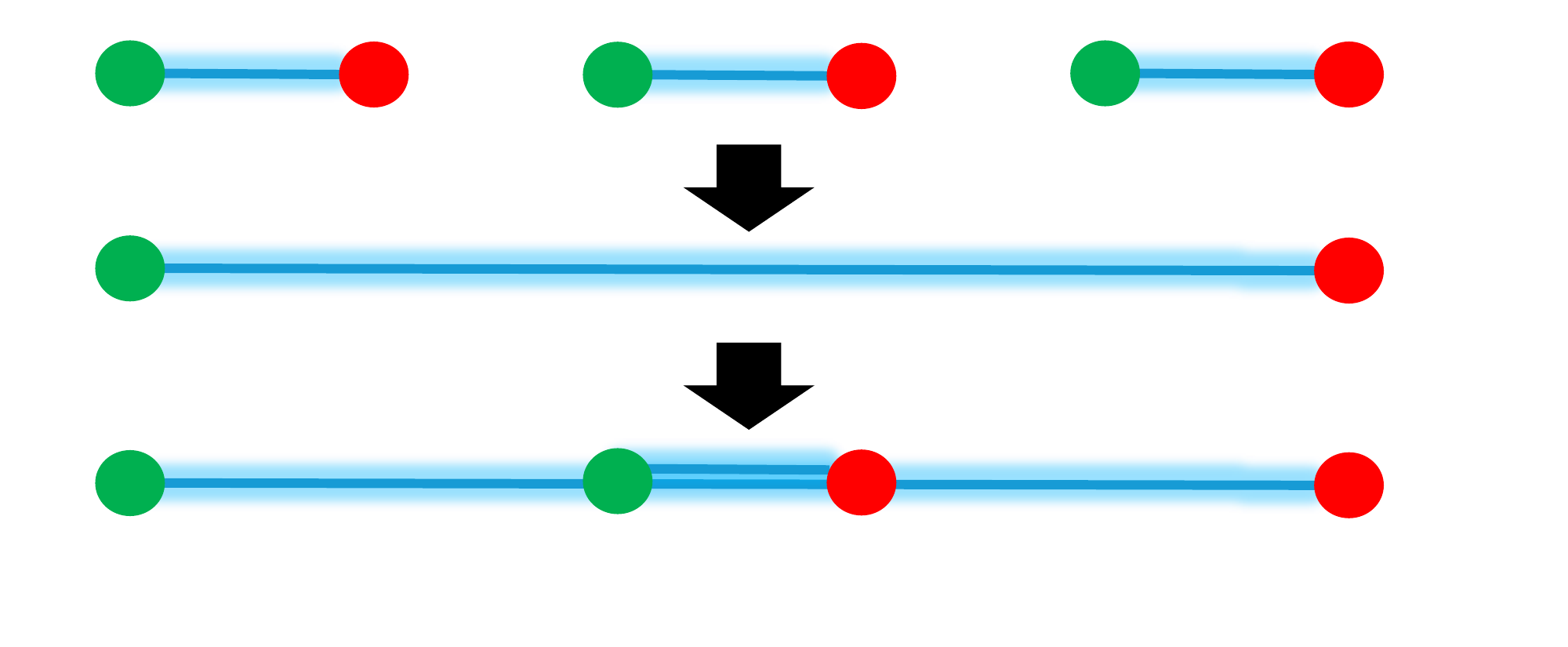}
    \caption{One possible way of increasing electric field strength by $2$. The green circles correspond to positive charges, and the red circles correspond to negative charges. First, the hopping operators are applied to every other link to create three $q\bar{q}$ pairs. The other hopping operators are then applied to remove the $q\bar{q}$ pairs in the center of the lattice. In the last step, a $q\bar{q}$ pair is excited in the center of the lattice.}
    \label{fig:SchwingerC2}
\end{figure}

Simulations of lattice QCD will require including quarks in addition to the gauge theory degrees of freedom. To understand how the inclusion of matter affects the truncation error, we will apply our framework to the Schwinger model. The Hamiltonian is given by
\begin{align}
    \hat{H} &= \hat{H}_K + \hat{H}_m + \hat{H}_E \nonumber \\
    \hat{H}_K &= \frac{1}{2} \sum_x \hat{\psi}^\dagger_{x+1} \hat{U}_x \hat{\psi}_x + \text{h.c.} \nonumber \\
    \hat{H}_m &= m\sum_x (-1)^x \hat{\psi}^\dagger_x \hat{\psi}_x \nonumber \\
    \hat{H}_E &= \frac{g^2}{2} \sum_x \hat{E}^2_x 
    \,,
\end{align}
where $\hat{\psi}_x$ is the fermion field at site $x$, $\hat{U}_x$ is the parallel transporter between sites $x$ and $x+1$, and $\hat{E}_x$ is the corresponding electric field. 
Note that in this theory, all gauge fields are sourced by fermions somewhere on the lattice. 
To have an electric field of strength $\Lambda$, there needs to be at least $\Lambda$ positively charged particles and $\Lambda$ negatively charged particles. 
This means that to raise the maximum electric field in a state by $\Delta$, $\Delta$ pairs of particles and anti-particles must be created. 
Furthermore,  one can have at most one fermion on each lattice site, such that all $\Lambda$ fermions need to sit on one side of the link with electric field $\Lambda$, while the anti-fermions sit on the other side. 
Starting from the electric vacuum, creating a single electric field requires a single hopping term. 
Creating an additional electric flux on this link requires first moving the fermion and anti-fermion created in the first step by two lattice units (because of staggering) to the left and right, requiring $2\times 2 = 4$ more hopping terms, and then adding an additional fermion anti-fermion pair. This requires a total of $6$ hopping terms.
Creating $3$ units of electric flux requires moving the two fermion anti-fermion pairs once more, requiring $4 \times 2 = 8$ before adding the extra pair, giving $15$ hopping terms.
In general, to create $n$ units of flux requires at least $n(2n-1)$ hopping terms.
This complicates the perturbative calculation of the leakage probabilities compared to the pure gauge case, but should also lead to a stronger suppression of states with large electric fields. 
To perform the perturbative estimate, the truncated Hamiltonian $\hat{H}_\Lambda$ is related to the untruncated Hamiltonian by 
\begin{align}
    \hat{H} &= \hat{H}_\Lambda + \sum_x \hat{V}_\Lambda^x \nonumber \\
    \hat{V}_\Lambda^x &= \frac{1}{2}\hat{\psi}^\dagger_{x+1} \hat{U}^\Lambda_x \hat{\psi}_x + \text{h.c.} \nonumber \\
    \hat{U}^\Lambda_x & = \sum_{n>\Lambda} \ket{n-1}\bra{n} + \sum_{n\leq-\Lambda} \ket{n-1}\bra{n} \, .
\end{align}
Following the same approach as the previous sections, the leading correction to the truncated time evolution is given by
\begin{equation}
        (e^{-i\hat{H}t} - e^{-i\hat{H}_\Lambda t})\ket{\phi(0)} = \sum_{x} - i e^{-i\hat{H}_\Lambda t} \int_{0}^{t} ds \ e^{i\hat{H}_\Lambda s} \hat{V}^{x}_\Lambda  e^{-i\hat{H}_\Lambda s} \ket{\phi(0)}
        \,.
\end{equation}
Assuming that $\Lambda$ is chosen to be the electric field strength where HSF begins, the leakage amplitude to go from an electric field of $\Lambda$ to $\Lambda+1$ can be estimated using a lattice with two staggered sites (one physical site), open boundary conditions, and an external electric field. For a system of this size, the gauge invariant states can be specified by basis states $\ket{E,f_0,f_1}$ where $E$ is an external electric field and $f_i$ are the fermion occupation numbers on each site. 
$f_0=1$ corresponds to a quark being present on site $0$ and $f_1=0$ corresponds to an anti-quark being present on site $1$. 
The leading order contribution to the truncation error is 
\begin{align}
(e^{-i\hat{H}t} - e^{-i\hat{H}_\Lambda t})\ket{\phi(0)} & = -i e^{-i t (\frac{g^2}{2}(\Lambda+1)^2 + 2m)}\ket{\Lambda+1,1,0} \bra{\Lambda+1,1,0} \int_0^t dt' e^{i\hat{H}_\Lambda t'}\hat{V}_\Lambda e^{-i\hat{H}_\Lambda t'}\ket{\phi(0)} \nonumber \\
&=\frac{ -i e^{-i t (\frac{g^2}{2}(\Lambda+1)^2 + 2m)}}{2} \ket{\Lambda+1,1,0} \int_0^t dt' c_{\Lambda,0,1} e^{i t(2m + \frac{g^2}{2}(2\Lambda+1))} \,
\end{align}
where $c_{\Lambda,0,1}$ is defined analogously to the previous sections.
This quantity is upper-bounded by
\begin{equation}
    \abs{(e^{-i\hat{H}t} - e^{-i\hat{H}_\Lambda t})\ket{\phi(0)}} \leq \frac{1}{2m+g^2(\Lambda + 1/2)} \, .
    \label{eq:SchwingerLeak1}
\end{equation}
Note that the amplitude to increase the electric field is suppressed when either $m$ is large or $g^2\Lambda$ is large. 
Increasing the electric field by two requires $6$ hopping terms acting on $6$ staggered sites. 
As before, the state of a lattice with $6$ staggered sites is given by an external electric field and the fermion occupation number on $6$ sites. 
Therefore, if HSF begins at $\Lambda-1$, the leading error in the state on a lattice with $6$ staggered sites is given by the expression
\begin{align}
    (e^{-i\hat{H}t} - &e^{-i\hat{H}_\Lambda t})\ket{\phi(0)} \nonumber\\
    &= 14(-i)^6 e^{-i\hat{H}_\Lambda t} \left(\prod_{k=1}^6 \int^{t_{k-1}}_0 dt_k \right) \hat{V}^3_{\Lambda}(t_1)\hat{V}^4_{\Lambda-1}(t_2)\hat{V}^2_{\Lambda-1}(t_3)\hat{V}^5_{\Lambda-1}(t_4)\hat{V}^3_{\Lambda-1}(t_5)\hat{V}^1_{\Lambda-1}(t_6) \ket{\phi(0)} \nonumber \\ 
    & =  -14 \ket{\Lambda-1,1,1,1,0,0,0}\left(\prod_{k=1}^6 \int^{t_{k-1}}_0 dt_k \right) 
    \nonumber\\
    & \qquad \qquad \times\bra{\Lambda-1,1,1,1,0,0,0}\hat{V}^3_{\Lambda}(t_1)\hat{V}^4_{\Lambda-1}(t_2)\hat{V}^2_{\Lambda-1}(t_3)\hat{V}^5_{\Lambda-1}(t_4)\hat{V}^3_{\Lambda-1}(t_5)\hat{V}^1_{\Lambda-1}(t_6) \ket{\phi(0)} \nonumber \\
    & =  -14 \ket{\Lambda-1,1,1,1,0,0,0}\left(\prod_{k=1}^6 \int^{t_{k-1}}_0 dt_k \right) 
   \frac{1}{2^6} e^{i t_1 (2m + 2g^2(\Lambda+1))} e^{i t_2 (-2m + g^2(\Lambda+\frac{1}{2}))}  \nonumber \\
   & \qquad \qquad \times 
   e^{i t_3 (-2m + g^2(\Lambda+\frac{1}{2}))}e^{i t_4 (2m + g^2(\Lambda-\frac{1}{2}))}e^{i t_5 (2m + g^2(\Lambda-\frac{1}{2}))} e^{i t_6 (2m + g^2(\Lambda-\frac{1}{2}))}c_{\Lambda-1,0,1,0,1,0,1} \nonumber \\
   & =A(\Lambda,g,m,t) c_{\Lambda-1,0,1,0,1,0,1} \ket{\Lambda-1,1,1,1,0,0,0}
   \label{eq:SchwingerLeak2}
\end{align}
where $\hat{V}^x_{\Lambda}(t) = e^{i\hat{H}_\Lambda t} \hat{V}^x_{\Lambda_0} e^{-i\hat{H}_\Lambda t}$, $t_0=t$, and $A(\Lambda,g,m,t)$ has been defined to be the integral in the above expression. The factor of $14$ comes from counting the number of possible orderings of operators that can be applied to raise the electric field on the center link by two. One possible ordering of operators to raise the field truncation is shown in Fig.~\ref{fig:SchwingerC2}. From this expression, it can be seen that the leading contribution to the error scales as $1/(g^2 \Lambda)^6$. This is a much more rapid falloff than in the equivalent scenario in the pure gauge theory. In high spatial dimensions with both gauge fields and matter, the electric field on a link can be raised by applying either plaquette or link operators. The leading terms in the truncation error will contain at most one hopping operator, and the rest will be plaquette operators. Consequently, the truncation errors in the $1+1D$ Schwinger model will not be representative of the behavior of truncation errors in higher spatial dimensions. However, the more rapid convergence of truncation errors in $1+1D$ suggests that non-trivial simulations of $1+1D$ dynamics with controlled uncertainties may be within near-term reach.

As in the pure gauge case, these estimates of leakage errors can be used to determine the truncation error of local observables. The error in a local observable will be given by the expectation of the observable in the state created by applying $\sum_x \hat{V}^x_\Lambda$ to the truncated state weighted by the probability of reaching that state. If the expression in Eq.~\eqref{eq:SchwingerLeak1} is used to estimate the leakage probability, the error in the electric energy ($\hat{E}_l^2$) and chiral condensate ($\hat{\chi}_x=(-1)^x\hat{\psi}^\dagger_x \hat{\psi}_x$) is given by
\begin{align}
    \delta E^2 &= (2\Lambda^2 + (\Lambda+1)^2)\times\left(\frac{1}{2m+g^2(\Lambda + 1/2)}\right)^2\nonumber \\
    \delta\chi &= 2\times\left(\frac{1}{2m+g^2(\Lambda + 1/2)}\right)^2
    \,.
    \label{eq:SchwingerErr1}
\end{align}
Note that the electric energy receives a contribution from the link itself being excited to $\Lambda+1$ and when its neighbor is excited to $\Lambda+1$. For higher truncations, HSF will occur at electric field strengths below $\Lambda$, and the expression in Eq.~\eqref{eq:SchwingerLeak2} can be used to estimate the error. Defining
\begin{equation}
    M(\Lambda,g,m,t) = \text{max}_{\tau < t}\abs{A(\Lambda,g,m,\tau)} \,,
\end{equation}
the error in the electric energy and chiral condensate is given by
\begin{align}
    \delta E^2 &= \left(2(\Lambda-1)^2 + 4\Lambda^2 + (\Lambda+1)^2\right) M^2(\Lambda,g,m,t)\nonumber \\
    \delta \chi &= 4 M^2(\Lambda,g,m,t) \,,
    \label{eq:SchwingerErr2}
\end{align}
where $A(\Lambda,g,m,\tau)$ is defined in Eq.~\eqref{eq:SchwingerLeak2}.

\begin{figure}
    \centering
    \includegraphics[width=\linewidth]{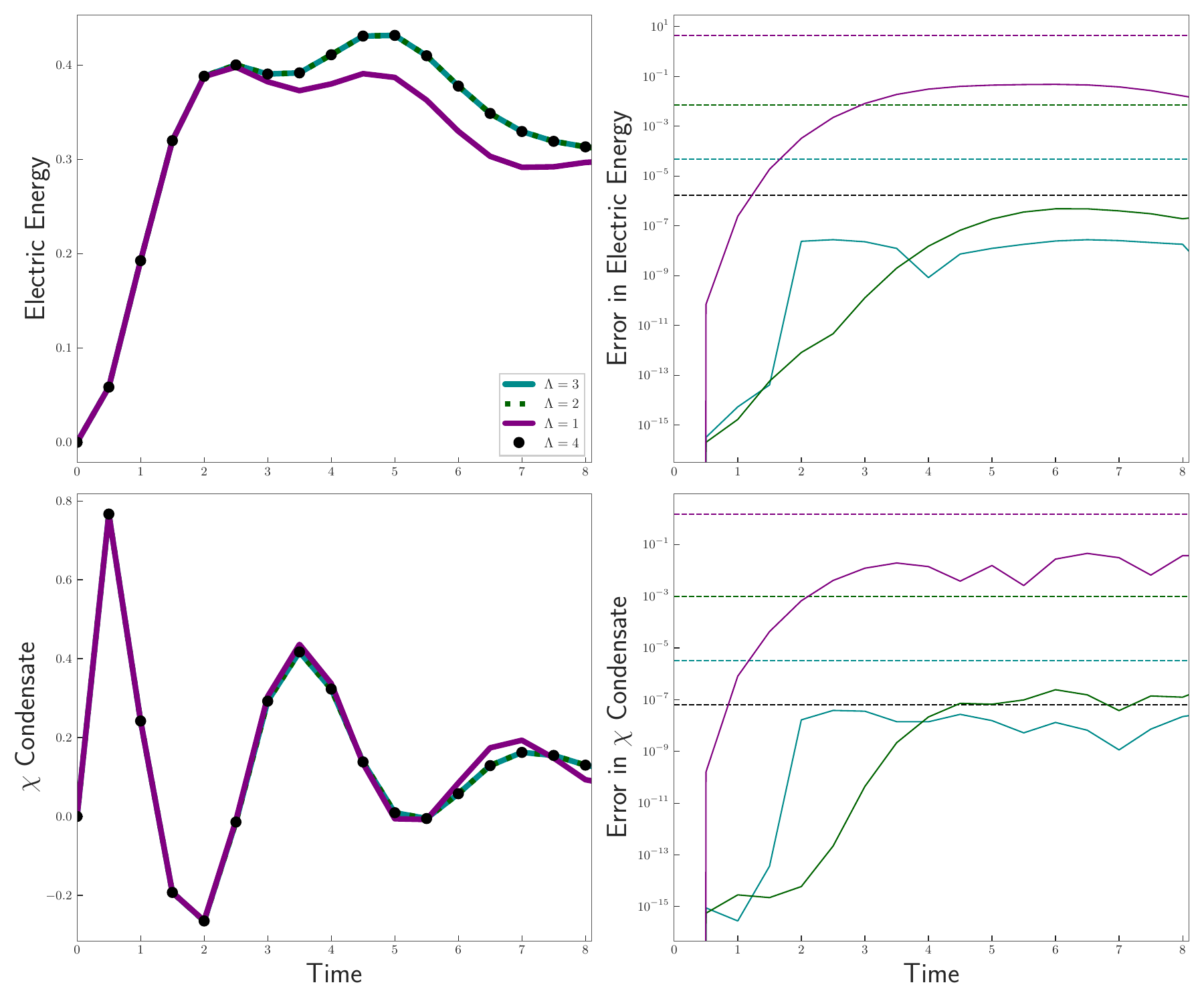}
    \caption{Evolution of the electric vacuum in the Schwinger model with $g=0.8$ and $m=0.1$. The left panel shows the electric energy per link and the chiral condensate per site as a function of time for different truncations. The right panel shows the difference in the expectation of the observable between the $\Lambda=4$ and lower truncations. The dashed lines in the right panel show the predicted truncation error. For $\Lambda=1$, the truncation error was computed using Eq.~\eqref{eq:SchwingerErr1} and for larger $\Lambda$, Eq.~\eqref{eq:SchwingerErr2} was used.}
    \label{fig:SchwingerSimulation}
\end{figure}

To demonstrate the performance of these error estimates, the dynamics of the Schwinger model were simulated using infinite MPS with $g=0.8$ and $m=0.1$. The details of the implementation are in Appendix~\ref{app:iMPS}. The system was initialized in the electric vacuum state and evolved in time until $t=8$. The left panel of Fig.~\ref{fig:SchwingerSimulation} shows the expectation of the electric energy and chiral condensate as a function of time for different truncations of the electric field. The truncation error from these numerical simulations is consistent with the analytical estimates in Eq.~\eqref{eq:SchwingerErr1} and Eq.~\eqref{eq:SchwingerErr2}.

\FloatBarrier
\section{Discussion}
In this work, a formalism has been developed for estimating the errors due to electric basis truncations in the simulation of lattice gauge theories. This formalism is based on the presence of Hilbert space fragmentation in the Kogut--Susskind Hamiltonian. This enables the use of time-dependent perturbation theory to describe the dynamics of states with large electric fields. Consequently, the leading contributions to the truncation error fall off factorially with the field truncation. Numerical simulations were performed to demonstrate the performance of these error estimates. The details of the truncation error depend on the change in electric energy from applying the plaquette operator.  This suggests lattices with different numbers of links per plaquette, such as a honeycomb lattice~\cite{Muller:2023nnk,Yao:2023gnm,Illa:2025dou,Illa:2025njz} or a triamond lattice~\cite{Kavaki:2024ijd,Kavaki:2025hcu}, will produce different truncation errors. As a result, there may exist an optimal choice that yields a stronger suppression of these errors.

The truncation errors estimated in this work are significantly smaller than in previous work. This is due to previous work being based on a loose estimate of the remainder term in the Dyson series expansion, while this work was able to derive the form of the dominant error due to truncation. As a result, for some parameter choices, our results reduce the error estimate by an astronomical factor of $10^{306}$. Our formalism does not apply to generic bosonic simulations; however, preliminary simulations in Appendix~\ref{app:Hubbard} suggest that there may be other mechanisms that restrict the growth of boson numbers more generally. While the formalism was only applied to Abelian gauge theories in one dimension in this work, the generalization to non-Abelian gauge groups and higher spatial dimensions is clear. This formalism also applies to scalar field theories with a compact scalar field, such as the $O(3)$ non-linear $\sigma$ model. The calculations performed in this work show that different observables have different dependencies on the truncation errors. One could potentially use this to construct observables with less sensitivity to truncation errors and potentially extract useful information from simulations with low truncations. This formalism can also be extended to improvement schemes using the similarity renormalization group to reduce the effects of truncation errors~\cite{Ciavarella:2023mfc}. These extensions of this work will enable quantum simulations of lattice QCD with complete theoretical control over the uncertainties due to truncating the gauge fields. 

\section{Author Contributions}
All authors contributed to the derivations and writing of the manuscript. A.C. and S.H. carried out the numerical simulations in this work.

\footnotesize
\begin{acknowledgments}
We would like to acknowledge useful conversations with Jesse Stryker, Ivan Burbano, Irian D'Andrea, Neel Modi, Chris Kane, and helpful feedback from John Preskill. A.N.C, S.H. and C.W.B. acknowledge funding through the U.S. Department of Energy (DOE), Office of Science under contract DE-AC02-05CH11231, partially through Quantum Information Science Enabled Discovery (QuantISED) for High Energy Physics (KA2401032). Additional support is acknowledged from the U.S. Department of Energy, Office of Science, National Quantum Information Science Research Centers, Quantum Systems Accelerator. J.C.H.~acknowledges funding by the Max Planck Society, the Deutsche Forschungsgemeinschaft (DFG, German Research Foundation) under Germany’s Excellence Strategy – EXC-2111 – 390814868, and the European Research Council (ERC) under the European Union’s Horizon Europe research and innovation program (Grant Agreement No.~101165667)—ERC Starting Grant QuSiGauge. Views and opinions expressed are, however, those of the author(s) only and do not necessarily reflect those of the European Union or the European Research Council Executive Agency. Neither the European Union nor the granting authority can be held responsible for them. This work is part of the Quantum Computing for High-Energy Physics (QC4HEP) working group.
\end{acknowledgments}
\normalsize

\appendix
\FloatBarrier
\section{Infinite MPS implementation}
\label{app:iMPS}
The simulations of the infinite plaquette chain were performed using the time-evolved block decimation algorithm (TEBD)~\cite{Vidal:2003lvx}. A Trotter step size of $\Delta t = 0.01$ was used. For the results shown in \cref{fig:plaq-infinite}, the time evolution was performed with a fixed bond dimension. The bond dimension was then increased by $10$, and the full simulation was rerun. This process was repeated until the difference in the electric energy between two consecutive runs was an order of magnitude smaller than the predicted truncation error. The resulting bond dimensions are shown in Table~\ref{tab:PlaqChainBondDim}. The results shown in \cref{fig:u1_prob} were obtained by similarly performing the evolution with a fixed bond dimension and then again after increasing the bond dimension by $25$. We repeated this process until the difference in the results from consecutive runs was 2 orders of magnitude below the upper bound computed using \cref{eq:prob_upper_bound} (corresponding to the dashed horizontal lines shown in \cref{fig:u1_prob}). We note that the simulation for all values of $g$ and $\Lambda$ converged within a bond dimension of $180$.  

\begin{table}[]
    \centering
    \begin{tabular}{|c|c|}
    \hline
        $\Lambda$ &  Bond Dimension \\
    \hline
        1 & 60 \\
    \hline
        2 & 90 \\
    \hline
        3 & 100 \\
    \hline
        4 & 170 \\
    \hline
        5 & 170 \\
    \hline
    \end{tabular}
    \caption{Bond dimension required for convergence of the electric energy for an infinite plaquette chain with $g=0.8$ and truncation $\Lambda$ shown in \cref{fig:plaq-infinite}.}
    \label{tab:PlaqChainBondDim}
\end{table}

The simulations of the Schwinger model were also performed using TEBD. The Hilbert space for the fermion on each site and the right neighboring link were combined into a single site, represented by a single tensor index in the matrix product state representation.  A Trotter step size of $\Delta t = 0.01$ was used. It was found that performing simulations with a large fixed bond dimension led to numerical instabilities. For this reason, the simulation was performed using a bond dimension of $50$ until $t=1.5$. The bond dimension was then increased, and the same criteria was used as in the plaquette chain simulation to determine when the simulation had converged to sufficient precision. The resulting maximum bond dimensions are shown in Table~\ref{tab:SchwingerBondDim}.

\begin{table}[]
    \centering
    \begin{tabular}{|c|c|}
    \hline
        $\Lambda$ &  Bond Dimension \\
    \hline
        1 & 90 \\
    \hline
        2 & 90 \\
    \hline
        3 & 110 \\
    \hline
        4 & 110 \\
    \hline
    \end{tabular}
    \caption{Bond dimension required for convergence of the electric energy for the Schwinger model on an infinite lattice with $g=0.8$, $m=0.1$, and truncation $\Lambda$.}
    \label{tab:SchwingerBondDim}
\end{table}

\FloatBarrier
\section{Hubbard-Holstein Model}
\label{app:Hubbard}
\begin{figure}
    \centering
    \includegraphics[width=\linewidth]{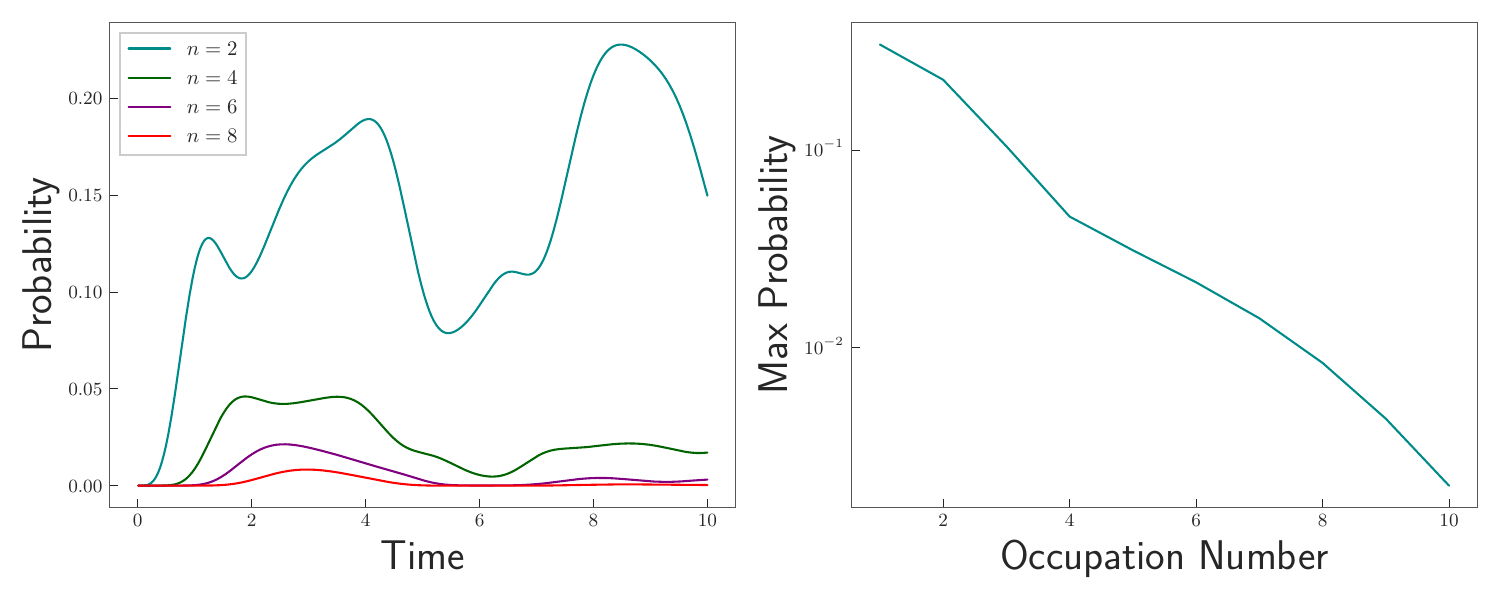}
    \caption{Simulation of the Hubbard-Holstein model on two sites with $g=1$ and $\omega=1$ for a time $t=10$. The left panel shows the probability of the boson mode on the left site having an occupation number of $n$ as a function of time. The right panel shows the maximum probability of occupying each boson number during this time evolution.}
    \label{fig:HubbardLeak}
\end{figure}

Previous work studying gauge field truncation errors for quantum simulation of lattice gauge theories also applied their results to bosonic systems~\cite{Tong:2021rfv}. The Hubbard-Holstein model was used as an example system. This theory is a model of electron-phonon interactions. The Hamiltonian is given by
\begin{align}
\hat{H} &= \hat{H}_F + g\hat{H}_{BF} + \omega \hat{H}_B \nonumber \\
\hat{H}_F &= \frac{1}{2} \sum_i \hat{\psi}^\dagger_{i+1} \hat{\psi}_i + \hat{\psi}^\dagger_{i} \hat{\psi}_{i+1} \nonumber \\
\hat{H}_{BF} &= \sum_i \hat{\psi}^\dagger_i \hat{\psi}_i (\hat{a}_i + \hat{a}^\dagger_i) \nonumber \\
\hat{H}_{F} &= \sum_i \hat{a}_i^\dagger \hat{a}_i \nonumber \\
\end{align}
where $\hat{\psi}_i$ is a fermion annihilation operator at site $i$, $\hat{a}_i$ is a boson annihilation operator at site $i$, $\omega$ is the phonon frequency and $g$ is the electron-phonon coupling. Note that this theory does not exhibit the same Hilbert space fragmentation that was used to derive truncation error estimates for lattice gauge theories. It was estimated that to simulate the dynamics of this model with $g=1$ for a time $t=1$, a boson truncation of $100$ was necessary to keep leakage errors below $\epsilon=0.2$. As a probe of this estimate, a simulation was performed of this model on two sites with $g=1$, $\omega=1$, and a single fermion present. In this simulation, the boson number at each site was truncated at $\Lambda=100$. The system was initialized with the boson number at both sites equal to zero, and the fermion was placed on the left site. The system was evolved for a time $t=10$. The left panel of Figure~\ref{fig:HubbardLeak} shows the probability of the boson mode on the left site having an occupation number of $n$ as a function of time. The right panel shows the maximum probability of occupying each boson number during this time evolution. As this figure shows, the leakage probability is significantly smaller than previous work estimated.

\bibliographystyle{quantum}
\bibliography{ref}

\end{document}